\documentclass{aa}

\usepackage[mathscr]{euscript}
\usepackage{times}
\usepackage{txfonts}
\usepackage{xcolor}
\usepackage{dirtytalk}
\usepackage{hyperref}
\newcommand{\orcid}[1]{\href{https://orcid.org/#1}{\includegraphics[width=8pt]{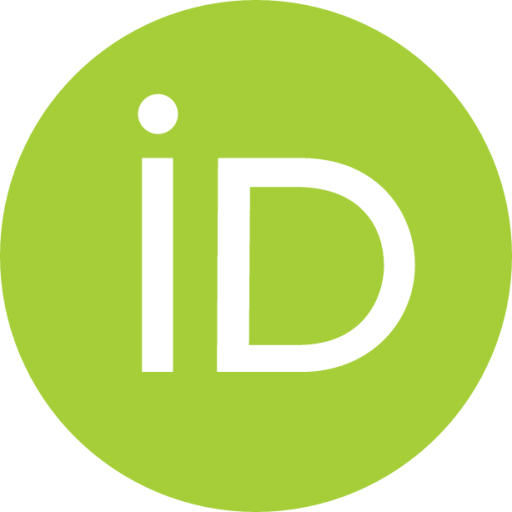}}}

\providecommand{\ut}{u^{t}} 
\providecommand{\uphi}{u^{\varphi}}

\providecommand{\sound}{c_{\mathrm{s},0}}
\providecommand{\cusp}{\beta_{\mathrm{cusp}}}
\providecommand{\Reppic}{\omega_{r}}
\providecommand{\Teppic}{\omega_{\theta}}
\providecommand{\xbar}{\bar{x}}
\providecommand{\ybar}{\bar{y}}

\providecommand{\en}{\mathscr{E}}

\providecommand{\aaa}{\mathscr{A}}
\providecommand{\nuR}{\nu_{\mathrm{R}}} \providecommand{\nuV}{\nu_{\mathrm{V}}} 
\providecommand{\nuRP}{\nu_{\mathrm{RP}}} 
\providecommand{\nuLT}{\nu_{\mathrm{LT}}}
\providecommand{\rg}{r_{\mathrm{G}}}

\graphicspath{{mat/}}

\usepackage{comment}

\begin{document}

\title{Accretion tori around rotating neutron stars}
\subtitle{Paper II: Oscillations and precessions } 
\titlerunning{Oscillations of tori around NSs}

\author{
   \orcid{0000-0002-4193-653X}Monika Matuszkov\'a\inst{\ref{inst1},\ref{inst2}}
    \and
    \orcid{0000-0003-3958-9441}Gabriel T\"{o}r\"{o}k\inst{\ref{inst1}
    }
    \and
    \orcid{0000-0002-0930-0961}Kate\v{r}ina Klimovi\v{c}ov\'a\inst{\ref{inst1}}
    \and
    \orcid{0000-0002-7635-4839}Ji\v{r}\'{\i} Hor\'{a}k\inst{\ref{inst2}}
    \and
    \orcid{0000-0001-5755-0677}Odele Straub\inst{\ref{inst3},\ref{inst4}}
    \and \\
    \orcid{0009-0000-7736-6180}Eva \v{S}r\'{a}mkov\'{a}\inst{\ref{inst1}}
    \and
    \orcid{0000-0003-0826-9787}Debora Lan\v{c}ov\'a\inst{\ref{inst1}}
    \and
    \orcid{0000-0001-9635-5495}Martin Urbanec\inst{\ref{inst1}}
    \and
    \orcid{0000-0002-4480-5914}Gabriela Urbancov\'a\inst{\ref{inst1}}
    \and
    \orcid{0000-0002-5760-0459}Vladim\'{\i}r Karas\inst{\ref{inst2}}
}
\institute{
    Research Centre for Computational Physics and Data Processing, Institute of Physics, Silesian University in Opava, 
    Bezru\v{c}ovo n\'am.~13, CZ-746\,01 Opava, Czech Republic
    \label{inst1}
    \and
    Astronomical Institute of the Czech Academy of Sciences, Bo\v{c}n\'{\i} II 1401, CZ-14100 Prague, Czech Republic \label{inst2}
    \and
    ORIGINS Excellence Cluster, Boltzmannstr. 2, 85748 Garching, Germany
    \label{inst3}
    \and Max Planck Institute for Extraterrestrial Physics, Gießenbachstra{\ss}e 1, 85748 Garching, Germany 
    \label{inst4}
}
          
\authorrunning{Matuszkov\'a et al.}

\date{Received / Accepted}

\abstract{
The four characteristic oscillation frequencies of accretion flows are, in addition to the Keplerian orbital frequency, often discussed in the context of the time variability of the black hole and neutron star (NS) low-mass X-ray binaries (LMXBs). These are namely the frequencies of the axisymmetric radial and vertical epicyclic oscillations, and the frequencies of non-axisymmetric oscillations corresponding to the periastron (radial) and Lense-Thirring (vertical) precessions. In this context, we investigate the effect of the quadrupole moment of a slowly rotating NS and provide complete formulae for calculating these oscillation and precession frequencies, as well as their convenient approximations. Simple formulae corresponding to the geodesic limit of a slender torus (and test particle motion) and the limit of a marginally overflowing torus (torus exhibiting a critical cusp) are presented, and furthermore, more general approximate formulae are included to allow calculations for arbitrarily thick tori. We provide the Wolfram Mathematica code used for our calculations together with {\texttt C++} and {\texttt{PYTHON}} codes for calculations of the frequencies. Our formulae can be used for various calculations describing the astrophysical signatures of the NSs’ superdense matter equation of state. For instance, we demonstrate that, even for a given fixed number of free parameters, a model accounting for fluid flow precession better matches the frequencies of twin-peak quasiperiodic oscillations observed in NS LMXBs than a model using geodesic precession.
} 


\maketitle

\section{Introduction}
\label{sec:intro}

Accretion discs are responsible for the strong observable radiation emitted by systems of accreting black holes (BHs) or neutron stars (NSs). Analysing their spectral and timing properties poses an excellent opportunity to explore strong gravity \citep{2002apa..book.....F,2008bhad.book.....K} as well as super-dense matter equation of state \citep[EoS,~][]{2007PhR...442..109L}. Here, we are especially motivated by several specific challenges arising in contemporary astrophysical research in the field of X-ray variability.

Throughout the history of X-ray timing observations of low-mass X-ray binaries (LMXBs) of both BH and NS types, numerous intriguing features have been revealed. The most rapid X-ray variability observed in these sources occurs at frequencies up to hundreds of Hertz, with the highest values exceeding $1.2$ kHz \citep[][]{2006MNRAS.370.1140B,2006csxs.book...39V}. Even though the rapid variability was discovered almost half a century ago, the origin of these so-called high-frequency (HF) quasi-periodic oscillations (QPOs), along with the correlated low-frequency (LF) QPOs, has never been sufficiently explained. Over the years, various models of QPOs have been proposed in an attempt to comprehend the nature of the phenomena \citep[see, e.g.,][ and references therein]{1985Natur.316..239A,1985Natur.317..681L,1998ApJ...508..791M,1999ApJ...520..763P,1999ApJ...524L..63S,2001A&A...374L..19A,2001PASJ...53....1K,2001AcPPB..32.3605K,2001ApJ...559L..25W,2002ApJ...577L..23T,2003PASJ...55..467A,2003MNRAS.344..978R,2004ApJ...603L..89K,2004A&A...423..401Z,2005AN....326..849B,2005A&A...439L..27P,2008A&A...487..527C,2008MNRAS.391.1332W,2009ApJ...694..387M,2010MNRAS.403.1193B,2011MNRAS.412.1659D,2014PhRvD..89f5007S,2016MNRAS.457.3859H,2016ApJ...819..112L,2017PhRvD..96j3015G,2020Univ....6...26S,2020MNRAS.497.2893W,2021ApJ...906...92S,2022ApJ...929...28T}. Most hypotheses link the QPO phenomenon to the orbital motion since the observed QPO frequencies are of the same order as those associated with the orbital motion in the innermost region of accretion discs.

Numerous studies have focused on the possible relation between the QPOs and oscillatory motion of stationary pressure-supported fluid configurations (accretion tori) formed near the central compact object \citep[e.g.,][]{2001astro.ph..5057K, 2004ApJ...603L..89K, 2003A&A...404L..21A,2003PASJ...55..467A,2003MNRAS.344L..37R,2003MNRAS.344..978R,2005AN....326..849B,2005A&A...436....1T,2011MNRAS.412.1659D,2016MNRAS.457L..19T,2018MNRAS.474.3967D,2023MNRAS.525.1126F}.  In this context, \citet{2009CQGra..26e5011S} and \citet{2016MNRAS.461.1356F} examined oscillation modes of fluid tori in the Kerr geometry, which describes spacetime around a rotating BH. In their investigation, in addition to other calculations, they evaluated the frequencies of these modes. Following their approach, we consider slowly rotating NSs for which the surrounding spacetime can be represented by the Hartle-Thorne geometry  \citep{1967ApJ...150.1005H,1968ApJ...153..807H}. This spacetime geometry is utilised in \cite{paperI} (hereafter Paper I), where, for the first time, a comprehensive analytical description of accretion tori around rotating NSs is provided.

In this paper, we focus on the epicyclic oscillation modes and their frequencies, especially on the axisymmetric and first-order non-axisymmetric modes. The first pair of modes corresponds to a simple radial and vertical motion of the torus. In the limit of a slender (zero thickness) torus, these modes' frequencies match the radial and vertical epicyclic frequencies of the geodesic motion of a test particle.

Similarly, the second pair of modes (which can be referred to as the precession modes) correspond to the torus' non-axisymmetric radial and vertical motion, and in the limit of a~slender torus, these modes' frequencies match the geodesic periastron and the Lense-Thirring precession frequencies respectively.

Our paper is organised as follows. In Sect.~\ref{Section:tori}, we briefly recall the properties of accretion tori around rotating NSs studied in Paper I. Derivation of the formulae determining these tori's oscillation modes is the primary subject of Sects.~\ref{Section:oscillations} and \ref{Section:perturb}. From Sect.~\ref{section:epicyclic}, special attention is given to investigating the behaviour of frequencies of the axisymmetric and precession modes. A complete set of formulae for calculating the frequencies of the epicyclic modes under consideration is provided within \textit{Wolfram Mathematica} code and additional \textit{C++} and {\textit{PYTHON}} libraries. In Sect.~\ref{Section:results}, we analyse the behaviour of these frequencies with a focus on the astrophysically relevant range of the NS parameters.
We also provide simplified versions of the formulae for the frequencies of axisymmetric radial and vertical oscillations and the frequencies of the two precession modes in Sect.~\ref{Section:formulae}. Finally, in Sect.~\ref{Section:conclusions}, we demonstrate the applications of our formulae on various astrophysically relevant examples, provide the reader with a short summary and state our main concluding remarks.

\section{Accretion tori around rotating neutron stars}
\label{Section:tori}

In Paper I, we describe a stationary non-self-gravitating 
fluid torus orbiting around a slowly rotating NS.\footnote{We use units where $c = G = 1$ with $c$ being the speed of light and $G$ the gravitational constant. To measure distance, we use $\rg = GM/c^2$, and  the $(-\,+\,+\,+)$ metric signature is employed. Finally, the $M_{\odot}$ symbol is used for the solar mass.} The fluid is described by the redshift function $A \equiv \ut$ and its orbital velocity with respect to a distant observer, which can be written as $\Omega \equiv \uphi/\ut$. Specific angular momentum of the fluid $l$ is assumed to be constant throughout the flow ($l = l_{0}$) and the specific angular momentum of the material is Keplerian at the torus centre at radius $r_{0}$,
$l_0=l_\mathrm{K}(r_0)$.
Furthermore, the fluid is assumed to be polytropic with rest-mass density $\rho$, pressure $p$, total energy density $e$, and polytropic constant $n$.

While the results of Paper I are applicable for a wide range of polytropic indices $0<n<\infty$ and a particular value of $n$ influences only the \say{labelling} of the equipressure surfaces, leaving the overall shape of the torus unaffected, the oscillation properties of the torus are sensitive to $n$. Here, in Paper II, we restrict ourselves almost exclusively to the value $n=3$, which corresponds to the radiation-pressure-dominated fluid at the innermost parts of the accretion flow. Finally, for comparison, we also present results for $n$ between $n=3/2$, corresponding to mono-atomic flows dominated by gas pressure, up to the values $n\gtrsim3/2$ corresponding to a more efficient heat exchange between different parts of the fluid by radiation.

The torus fills up closed equipotential surfaces given by the condition 
\begin{equation}
    f\left( r, \theta \right) = 1 - \frac{1}{n\sound^2}  \left( 1 -   \frac{\en_0}{\en} \right) = \mathrm{const.},
    \label{f}
\end{equation}
where $\en = - u_{t}$ is specific mechanical energy.

The radial and vertical torus proportions are determined by the thickness parameter $\beta$, which is directly proportional to the sound speed in the torus centre $\sound$
\begin{align}
	\beta^2 =& \frac{2n\sound^2}{r_0^2 \Omega_0^2 A_0^2}.
\end{align}

When $r_{0}$ is small enough, a critical value of the $\beta$ parameter exists ($\beta=\cusp$). This corresponds to the largest closed equipotential determining the torus surface with a cusp in the equatorial plane. On the other hand, in the slender-torus limit ($\beta \rightarrow 0$), the poloidal cross-section has an elliptical shape with semi-axes in the ratio of the radial ($\Reppic$) and vertical ($\Teppic$) epicyclic frequencies of a free test particle at $r_0$ as measured by a distant observer
\begin{equation}
    f^{(0)} = 1 - \bar{\omega}_r^2\bar{x}^2 - \bar{\omega}_\theta^2\bar{y}^2 = \mathrm{const.},
    \label{f0}
\end{equation}
where $\xbar$, $\ybar$ are  coordinates scaled by the $\beta$ parameter
\begin{align}
	\xbar =& \frac{\sqrt{ g_{rr,0} }}{\beta} \left( \frac{r - r_0}{r_0} \right)
 \\ \mathrm{and}\nonumber\\ 
	\ybar =& \frac{\sqrt{g_{\theta\theta,0}}}{\beta} \left( \frac{\frac{\pi}{2} - \theta}{r_0} \right).
\end{align}

The NS spacetime is represented by the Hartle-Thorne geometry characterised by three parameters: NS mass $M$, dimensionless angular momentum $j$ and dimensionless quadrupole moment $q$. A detailed description of the metric, the solution for the tori, and the study of the influence of the metric parameters on the shape, size and structure of the torus are presented in Paper~I. Within that paper, we also provide a Wolfram Mathematica notebook containing the full set of equations for the torus. In the following, we examine the oscillation modes of these tori.

\section{Oscillations}
\label{Section:oscillations}

The description of oscillations of tori can be inferred using a linear perturbation of the equations describing the relativistic conservation laws. Since the equilibrium tori are stationary and axially symmetric, Eulerian perturbations of all quantities can be found in the form of normal modes
\begin{equation}
    \delta X (t, r, \theta, \varphi) = \delta X (r, \theta)\, \mathrm{e}^{i \left( m \varphi - \omega t \right)},
\end{equation}
where $m$ denotes azimuthal wavenumber and $\omega$ is the oscillation frequency with respect to the coordinate time. Its value, as well as the shape of the perturbation in the poloidal plane $\delta X(r,\theta$), follow from the solution of the eigenvalue problem in two dimensions
\begin{equation}
    \hat{L}W + 2n\left(\bar{\omega} - m\bar{\Omega}\right)^2 W = 0
    \label{papa}
\end{equation}
with
\begin{align}
    \hat{L} = \frac{\beta^2 r_0^2}{\aaa^2}\left[
    \frac{1}{\sqrt{-g}f^{n-1}}\frac{\partial}{\partial x^j}
    \left(\sqrt{-g}g^{jk}\alpha f^n \frac{\partial}{\partial x^k}\right) \right.
    \nonumber \\
    \left. - \frac{g^{\varphi\varphi} - \Omega g^{t\varphi}}{1-l\Omega}
    \left(m-l\omega\right)^2\alpha f\right],
    \label{papa-L}
\end{align}
where $ \lbrace j,k \rbrace \in \lbrace r, \theta \rbrace $ are the poloidal coordinates, $ \aaa \equiv A/A_0 $, 
$ \bar{\Omega} \equiv \Omega / \Omega_0 $,   
$ \bar{\omega} \equiv \omega / \Omega_0 $,
and $g$ is the determinant of the metric tensor. The normalised frequency of the oscillations $\bar{\omega}$ is the eigenvalue of the problem. The eigenfunction $W$ is related to the pressure perturbation $\delta p$ as~\citep{2006CQGra..23.1689A}
\begin{equation}
    W = - \frac{\delta p}{A \rho \left( \omega - m \Omega \right)},
\end{equation}
and it has to satisfy
\begin{equation}
    \lim_{f\rightarrow 0}\left(f^n W\right) = 0
\end{equation}
at the surface of the torus. The last equation simply expresses the free-surface boundary condition. Once a particular mode of oscillation is found, the pressure and the poloidal velocity perturbations result from
\begin{equation}
    \delta p = -\rho\left(\omega - m\Omega\right)AW,
    \quad
    \delta u_k = \frac{i}{H}\frac{\partial W}{\partial x^k}, 
\end{equation}
where $H$ denotes the specific enthalpy. Finally, the quantity $\alpha$ in Eq.~(\ref{papa-L}) reflects a correction introduced by the relativistic equation of state. In all previous studies, the contribution of gas internal energy to the total mass density was neglected ($e\approx\rho$), corresponding to $\alpha=1$. Here, we use a consistent expression 
\begin{equation}
    \alpha = \frac{1}{1-n\sound^2\left(1-f\right)}.
\end{equation}

\subsection{Slender torus limit}
Equation (\ref{papa}) is a relativistic version of the Papaloizou-Pringle equation \citep{1984MNRAS.208..721P}. It has no analytical solution except for the limit case of an infinitely slender torus $ \left( \beta \rightarrow 0 \right)$. In this limit, Eq.~(\ref{papa}) becomes
\begin{equation}
    \hat{L}^{(0)}W^{(0)} + 2n\left(\bar{\omega^{(0)}}-m\right)^2 W^{(0)} = 0,
    \label{slender-papa}
\end{equation}
where
\begin{equation}
    \hat{L}^{(0)} = 
    \frac{1}{\left(f^{(0)}\right)^{n-1}}\left(
    \frac{\partial}{\partial\bar{x}}
    \left[\left(f^{(0)}\right)^n\frac{\partial}{\partial\bar{x}}\right] +
    \frac{\partial}{\partial\bar{y}}
    \left[\left(f^{(0)}\right)^n\frac{\partial}{\partial\bar{y}}\right]\right) 
    \label{slender-papa-L}
\end{equation}
is the leading-order approximation of $\hat{L}$ as $\beta\rightarrow 0$. It is worth noting that Eqs.~(\ref{slender-papa}) and (\ref{slender-papa-L}) do not depend on the spacetime metric and, moreover, are identical to those in Newtonian gravity. The only quantities that depend on the gravitational field are the normalised epicyclic frequencies in the $f^{(0)}$ function in Eq.~(\ref{f0}).

\subsection{Epicyclic modes in the slender torus limit}

As a result of the small radial and vertical perturbation of the motion of the free test particle, which initially moves along a circular geodesic path, the particle performs radial and vertical epicyclic oscillations. A natural generalisation of the concept of epicyclic oscillations of free test particles to fluid configuration leads to the epicyclic oscillation modes excited by nearly uniform displacements of the torus from its equilibrium.

\citet{2006CQGra..23.1689A} demonstrated that in the case of slender tori, the pressure force is insignificant, and the epicyclic oscillations of the individual fluid elements are identical to those of free particles. In the Eulerian framework, at a fixed azimuth, the frequencies of oscillations are simply $\omega_{r,m}^{(0)} =\omega_{r 0} + m\Omega_0$ and $\omega_{\theta,m}^{(0)} = \omega_{\theta 0} + m\Omega_0$, where $m$ is the integer azimuthal wavenumber (number of arms of the mode in the azimuthal direction). 
The axisymmetric modes, for which all the fluid components oscillate with the same phase, are associated with the value $m=0$. The value $m=-1$ then corresponds to eccentric and tilted configurations in the case of the radial and vertical modes, respectively. In Newtonian gravity, frequencies of this mode vanish because $\omega_{r0} = \omega_{\theta 0} = \Omega_0$, and both modes describe stationary perturbation to another equilibrium. However, as the frequencies start to deviate from each other, the $m=-1$ modes describe pericentric (apsidal) precession of the torus with the frequency $\omega_\mathrm{RP}=\Omega_0-\omega_{r0}$, or a precession of its orbital plane (``Lense-Thirring precession'') with frequency $\omega_\mathrm{LT}=\Omega_0-\omega_{\theta 0}$.

In the general relativistic framework, the former geodesic precession frequency is non-zero already in the spherically symmetric spacetime of a non-rotating star, but the latter arises only from the breaking of the spherical symmetry due to the star's rotation. It is therefore often considered as a possible sensitive measure of the BH or NS spin \citep{1999ApJ...524L..63S,1999ApJ...513..827M,2020A&A...643A..31K}.

\subsection{Non-slender tori}

\citet{2006MNRAS.369.1235B} showed that the eigenfunctions of the slender tori have polynomial dependence on $\bar{x}$ and $\bar{y}$ coordinates. The solution then can be easily found by comparing coefficients of the polynomials that appear on both sides of Eq.~(\ref{slender-papa}). The authors found all explicit expressions for the lowest order modes up to the cubic order. In particular, the epicyclic modes that are of interest here correspond to the linear eigenfunctions $W^{(0)}_r = \bar{x}$ and $W^{(0)}_\theta=\bar{y}$.

Later on, \citet{2007A&A...467..641S} and \citet{2009CQGra..26e5011S} inspired by the earlier work of \citet{Blaes1987} used the fact that the $\hat{L}$ operator in Eq.~(\ref{slender-papa}) is self-adjoint to derive the eigenfunctions and eigenfrequencies of the epicyclic modes in a non-slender torus case using a perturbative expansion in $\beta$ parameter. This method is also applied in this work for thicker tori in the Hartle-Thorne spacetime.

\section{Perturbative expansion in torus thickness}
\label{Section:perturb}

Since the exact solution is known for a simplified case (i.e. for $ \beta \rightarrow 0 $), we can use perturbation theory to find the solution for more complicated cases ($ \beta > 0 $). \footnote{Note that the perturbation method gives reasonable results only for small values of $\beta$ and our results are therefore valid only for slightly non-slender tori.}

By expanding the quantities $ \bar{\omega} $, $ W $, $\aaa$, $\bar{\Omega}$, $f $ in $ \beta $ \citep{2009CQGra..26e5011S}
\begin{equation}
	Q = Q^{(0)} + \beta Q^{(1)} + \beta^2 Q^{(2)} + \cdots, \qquad
	Q \in \left\lbrace \bar{\omega}, W, \aaa, \bar{\Omega}, f \right\rbrace,
	\label{rozvoj}
\end{equation}
substituting it into Eq.~(\ref{papa}) and comparing the coefficients of appropriate order in $ \beta $, we obtain the corresponding corrections to $W$ and $\omega$.

In the zeroth order, the problem reduces to the eigenvalue problem (\ref{slender-papa}). Its solution is represented by an infinite set of linear modes $\{\omega^{(0)}_A, W^{(0)}_A\}$. In particular, eigenfrequency and eigenfunction for the radial and vertical epicyclic modes are $\omega_{r,m}^{(0)} = \omega_{r0} + m\Omega_0$, $W_r^{(0)}=\bar{x}$, and $\omega_{\theta,m}^{(0)} = \omega_{\theta0} + m\Omega_0$ and $W_\theta^{(0)} = \bar{y}$, respectively. Since the $\hat{L}$ operator is self-adjoint with respect to the scalar product
\begin{equation}
    \left\langle W_1\,|\,W_2\right\rangle = \int_{S}W_1^\ast W_2 (f^{(0)})^{n-1}\mathrm{d}S,
\end{equation}
all zeroth-order eigenfunctions $W^{(0)}_A$ form an orthogonal complete set in the Hilbert space of smooth functions defined over the torus cross-section. This fact is of great use in the calculations of the higher-order corrections $W^{(i)}$ because they can be expanded in the basis of slender-torus eigenfunctions. Thus, the problem can be reduced to the solution of algebraic equations.

The equations governing the first-order corrections $W^{(1)}$ in the case of both epicyclic modes take the form
\begin{equation}
    \hat{L}^{(0)} W^{(1)} + 2n\left(\bar{\omega}^{(0)} - m\right)^2 W^{(1)} = \Phi^{(1)}
    \label{papa-1}
\end{equation}
with
\begin{equation}
    \Phi^{(1)} = -\hat{L}^{(1)} W^{(0)} - 
    4n\left(\bar{\omega}^{(1)} - m\bar{\Omega}^{(1)}\right)\left(\bar{\omega}^{(0)}-m\right) W^{(0)}.
\end{equation}
This equation differs from Eq.~(\ref{slender-papa}) only by the presence of $\Phi^{(1)}$ on the right-hand side. Setting $\Phi^{(1)}$ to zero essentially brings us to a zeroth-order problem described by Eq.~(\ref{slender-papa})) that has a nontrivial solution. Therefore, the right-hand side has to satisfy the solvability condition $\langle W^{(0)}\,|\,\Phi^{(1)}\rangle = 0$ to assure that the solution $W^{(1)}$ exists. The solvability condition immediately provides us with the first-order correction to the eigenfrequency
\begin{equation}
    \bar{\omega}^{(1)} = -\frac{\left\langle W^{(0)}\,|\,L^{(1)} W^{(0)}\right\rangle}{4n\left(\bar{\omega}^{(0)}-m\right)\left\langle W^{(0)}\,|\, W^{(0)}\right\rangle} - m\frac{\left\langle W^{(0)}\,|\,\bar{\Omega}^{(1)} W^{(0)}\right\rangle}{\left\langle W^{(0)}\,|\, W^{(0)}\right\rangle}.
\end{equation}
Similar to the tori in Kerr spacetime, the integration of the odd function over a symmetric domain here causes the scalar products in both numerators for both epicyclic modes to vanish \citep{2009CQGra..26e5011S}. Therefore, we may write $\bar{\omega}^{(1)} = 0$. The first-order correction $W^{(1)}$ can be found by solving Eq.~(\ref{papa-1}) with $-L^{(1)}W^{(0)} + 4nm \left(\bar{\omega}^{(0)} - m\right) \bar{\Omega}^{(1)} W^{(0)}$ on the right-hand side. As mentioned before, this is easily done by expanding the solution in the eigenfunctions of the zeroth-order problem and by projecting the equation on the individual eigenfunctions. We find 
\begin{equation}
    W^{(1)} = \sum_A c_A W^{(0)}_A,
    \label{W1}
\end{equation}
where
\begin{equation}
    c_A = \frac{\left\langle W^{(0)}_A\,|\, L^{(1)} W^{(0)}\right\rangle - 4nm \left(\bar{\omega}^{(0)} - m\right) \left\langle W^{(0)}_A\,|\, \bar{\Omega}^{(1)} W^{(0)}\right\rangle}{2n\left[
    \left(\bar{\omega}^{(0)}_A - m\right)^2 - \left(\bar{\omega}^{(0)} - m\right)^2\right]
    \left\langle W^{(0)}_A\,|\,W^{(0)}_A \right\rangle}.
    \label{cA}
\end{equation}
Since the eigenfunctions $W^{(0)}_A$ are polynomials, coefficients $c_A$ are nonzero only for a limited number of the participating modes. For the epicyclic modes, the $L^{(1)} W^{(0)}$ terms are quadratic even-parity polynomials in both $\bar{x}$ and $\bar{y}$. Consequently, the scalar product in the numerator of Eq.~(\ref{cA}) does not vanish only for modes whose eigenfunctions are at most quadratic in variables $\bar{x}$ and $\bar{y}$. According to the notation introduced by \citet{2006MNRAS.369.1235B}, these are the corotation mode, two epicyclic modes, `X', `plus', and breathing mode.  Thus, the sum in expression (\ref{W1}) only contains several terms.

Finally, the Papaloizou-Pringe Eq.~(\ref{papa}) expanded to the second order has a similar form as the previously discussed Eq.~(\ref{papa-1}),
\begin{equation}
    \hat{L}^{(0)} W^{(2)} + 2n\left(\bar{\omega}^{(0)} - m\right)^2 W^{(2)} = \Phi^{(2)}
    \label{papa-2}
\end{equation}
with
\begin{align}
    \Phi^{(2)} =& -\hat{L}^{(2)} W^{(0)} - \hat{L}^{(1)} W^{(1)} 
    \nonumber \\    
    &-2nm^2\Omega^{(1)2} W^{(0)} 
    + 4nm\Omega^{(1)}\left(\bar{\omega}^{(0)} - m\right) W^{(1)} 
    \nonumber \\
    &- 4n\left(\bar{\omega}^{(2)} - m\bar{\Omega}^{(2)}\right)
    \left(\bar{\omega}^{(0)} - m\right) W^{(0)}, 
\end{align}
where $\bar{\omega}^{(1)}=0$ is used. The solvability condition now gives the second-order correction to the eigenfrequency:
\begin{align}
    \bar{\omega}^{(2)} =& -\frac{\left\langle W^{(0)}\,|\, \hat{L}^{(1)} W^{(1)}\right\rangle +
    \left\langle W^{(0)} \,{\big|}\, 
    \left(L^{(2)} + 2nm^2\bar{\Omega}^{(1)2}\right) 
    W^{(0)}\right\rangle}
    {4n\left(\bar{\omega}^{(0)} - m\right)
    \left\langle W^{(0)} \,|\, W^{(0)}\right\rangle} 
    \nonumber \\
    &+ m\frac{\left\langle W^{(0)}\,|\,\bar{\Omega}^{(2)} W^{(0)}\right\rangle + 
    \left\langle W^{(0)}\,|\,\bar{\Omega}^{(1)} W^{(1)}\right\rangle}
    {\left\langle W^{(0)}\,|\, W^{(0)}\right\rangle}.
\end{align}
As both $L^{(2)}$ and $\Omega^{(2)}$ have even parity in $\bar{x}$ and $\bar{y}$, the integration in the scalar products is carried over the even functions, and, therefore, the scalar products do not in general vanish.

\section{Epicyclic and precession modes -- frequencies}
\label{section:epicyclic}

The procedure described in the previous Section can, in principle, be repeated for any oscillation mode up to an arbitrary order in the $\beta$ expansion. Next, we restrict our attention to the epicyclic modes and calculations reflecting the first non-trivial corrections to the slender-torus eigenfrequencies. As a result, approximately quadratic dependency on the torus thickness is obtained,
\begin{align}
    \label{epic}
	\omega_{i,m}\left(r_0, \beta\right) &= \omega_{i\,0} + m \, \Omega_{0} 
	+ \omega_{i,m}^{(2)}\,  \beta^2 + \mathscr{O} \left( \beta^3 \right)\,,
\end{align}
where the subscript $i\in \{r,\,\theta\}$ denotes the radial or vertical mode and $\omega_{i\,0}$
is the slender torus epicyclic frequency, which is equal to a test particle epicyclic frequency.

\subsection{Precessions}

For small eccentricities and inclinations of the torus, the pericentre (apsidal) and Lense-Thirring (nodal) precessions can be viewed as the $m=-1$ radial and vertical epicyclic modes. In the slender-torus limit, the eccentricity $e$ and tilt $\vartheta$ of the torus are related to the mode amplitudes $\mathcal{A}_r$ and $\mathcal{A}_\theta$ as follows
\begin{equation}
   e = \left(\frac{2}{A\omega_r r\sqrt{g_{rr}}}\right)_0 \mathcal{A}_r,
   \quad
   \vartheta = \left(\frac{2}{A\omega_\theta r^2\sqrt{g_{\theta\theta}}}\right)_0 \mathcal{A}_\theta.
\end{equation}
The eigenfunctions of the slender torus are assumed to be of the form $W_{r,-1} = \beta \mathcal{A}_r \bar{x} e^{-i\varphi}$ and $W_{\theta,-1} = \beta \mathcal{A}_\theta \bar{y} e^{-i\varphi}$. The eigenfrequencies are $\omega_{r,-1} = \omega_{r0} - \Omega_0$ and $\omega_{\theta,-1} = \omega_{\theta0} - \Omega_0$. Since Eqs.~(\ref{papa}) and (\ref{papa-L}) are invariant under the transformation $(\omega,m)\rightarrow(-\omega,-m)$, the precessions can be equally described by $m=1$ modes with frequencies $\omega_{r,1} = \Omega_0 - \omega_{r0}$ and $\omega_{\theta,-1} = \Omega_0 - \omega_{\theta0}$. In the case of thicker tori, the non-axisymmetric waves efficiently transport information between different radii so that the entire torus effectively responds as a single body with definite frequencies $\omega_{r,-1}$ and $\omega_{\theta,-1}$.

 \begin{figure*}[t]
	\begin{center}
		\includegraphics[width=\linewidth]{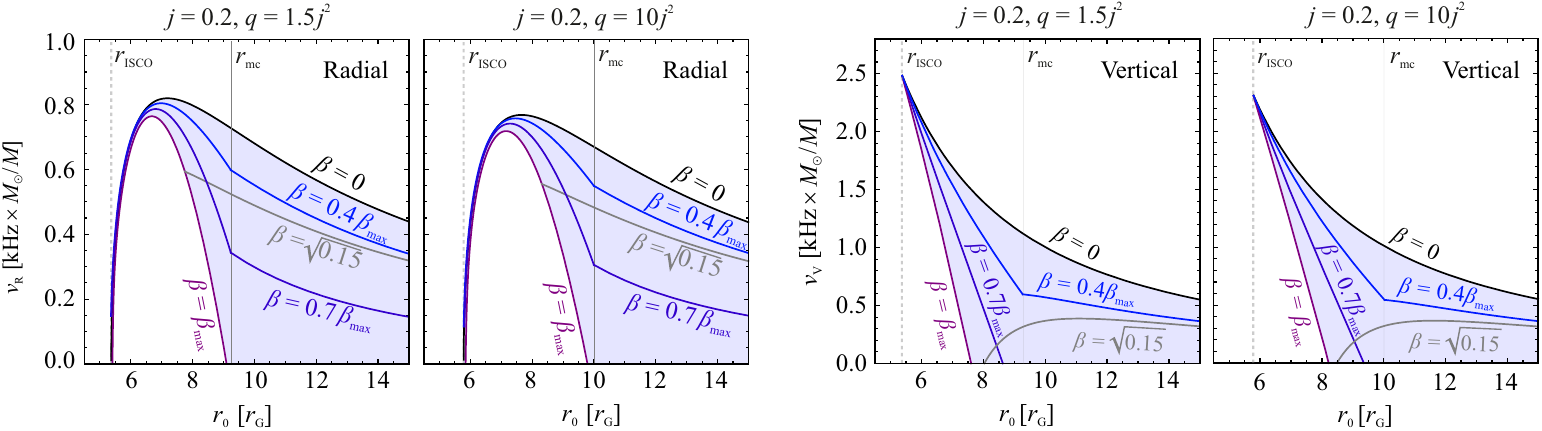}
		\caption{The axisymmetric  oscillation frequencies calculated for different $\beta$ and NSs with $j=0.2$, and two particular values of $q/j^2$, $q/j^2 = 1.5$ (extreme NS compactness), and $q/j^2 = 10$ (large NS oblateness). \emph{Left:} The radial mode. \emph{Right:} The vertical mode. The curves indicating particular values of $\beta$ are normalised to the maximal value $\beta=\beta_{\mathrm{max}}$. The critical radius limiting the existence of cusp tori is denoted by the dotted vertical line marked as $r_{\mathrm{mc}}$. The maximal value $\beta=\beta_{\mathrm{max}}$ matches the cusp value when $r<r_{\mathrm{mc}}$, and the value implying tori extending to infinity when $r>r_{\mathrm{mc}}$. For the absolute value of $\beta$, only one specific curve is drawn in each panel, marked in grey, $\beta=\sqrt{0.15}$, which makes it possible to distinguish frequencies that correspond to relatively large tori, as they are displayed below this curve. The dotted vertical lines in each panel marked as $r_{mc}$ denote the radii limiting the existence of cusp tori -- any torus with a centre above this radii does not exhibit the cusp. An analogous notation is later used in several other figures. \label{figure:frequencies:axi}
		}
	\end{center}
\end{figure*}

\subsection{Simplified notation and formulae in Mathematica}

Hereafter, we focus on the $m\in\{0,-1\}$ modes and use ordinary frequencies, expressed in the number of cycles per second, $\nu[\mathrm{Hz}]=\omega/2\pi$. For convenience, we introduce the simplified notation for the oscillatory modes' frequencies. Namely, for the axisymmetric modes, we set
\begin{align}
\nuR \equiv\nu_{r,0}\,,\quad
\nuV \equiv\nu_{\theta,0}\,,
\end{align}
and, for the precession modes, 
\begin{align}
\nuRP \equiv\nu_{r,-1}\,,\quad
\nuLT \equiv\nu_{\theta,-1}\,.
\end{align}

A complete set of formulae for calculating the frequencies of the four modes under consideration is provided within the Wolfram Mathematica code and additional {\textit C++} and {\textit{PYTHON}} libraries\footnote{\url{https://github.com/Astrocomp/HTtori_oscillations}}.

\section{Frequency behaviour}
\label{Section:results}

Assuming a linearised scenario, for a given spacetime geometry and torus centre $r_0$, the oscillatory mode frequency $\nu(\beta)$ should decrease or increase as the torus thickness parameter $beta$ increases. The value of $\nu(\beta_\mathrm{cusp})$ then represents a limit on $\nu$, as opposed to the geodesic case of $\nu(\beta\,=\,0)$. Such a simple behaviour was identified for several modes in the study of \citet{2009CQGra..26e5011S}, which took into account Kerr geometry and the influence of torus thickness up to the second order terms (i.e., $\beta^2$).

\begin{figure*}[t]
	\begin{center}
		\includegraphics[width=\linewidth]{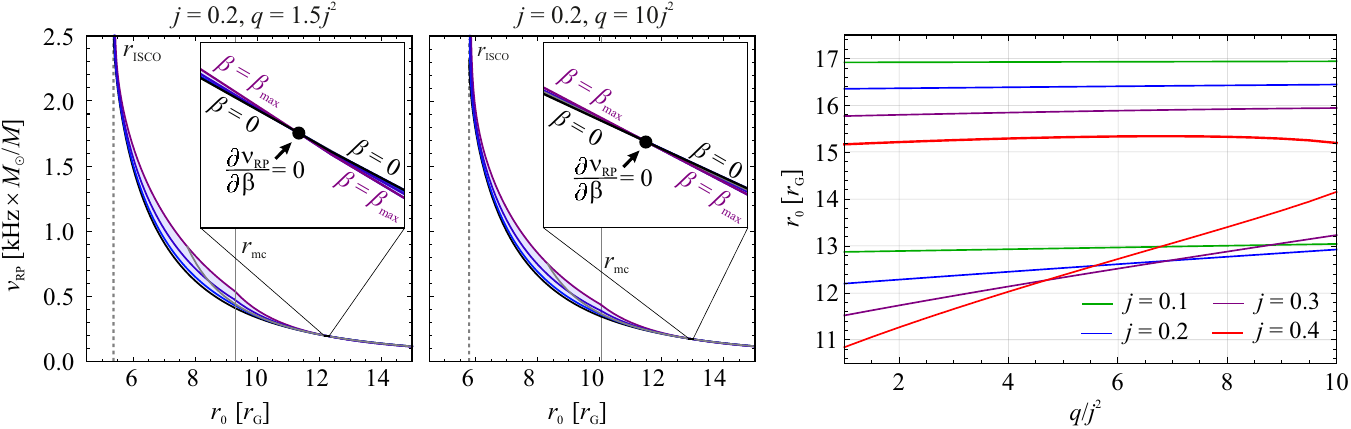}
		\caption{{\emph{Left:}} The radial precession frequencies calculated for different $\beta$ and NSs with $j=0.2$ and two particular values of $q/j^2$, $q/j^2 = 1.5$ (extreme NS compactness), and $q/j^2 = 10$ (large NS oblateness). {\emph{Right:}} Radial coordinates corresponding to the vanishing of the correction to the radial precession oscillation mode frequency, i.e., radii where $\nuRP^{(2)}=0$. 
		 \label{figure:r_prc}
  		}
	\end{center}
\end{figure*}

\begin{figure*}[t]
	\begin{center}
		\includegraphics[width=\linewidth]{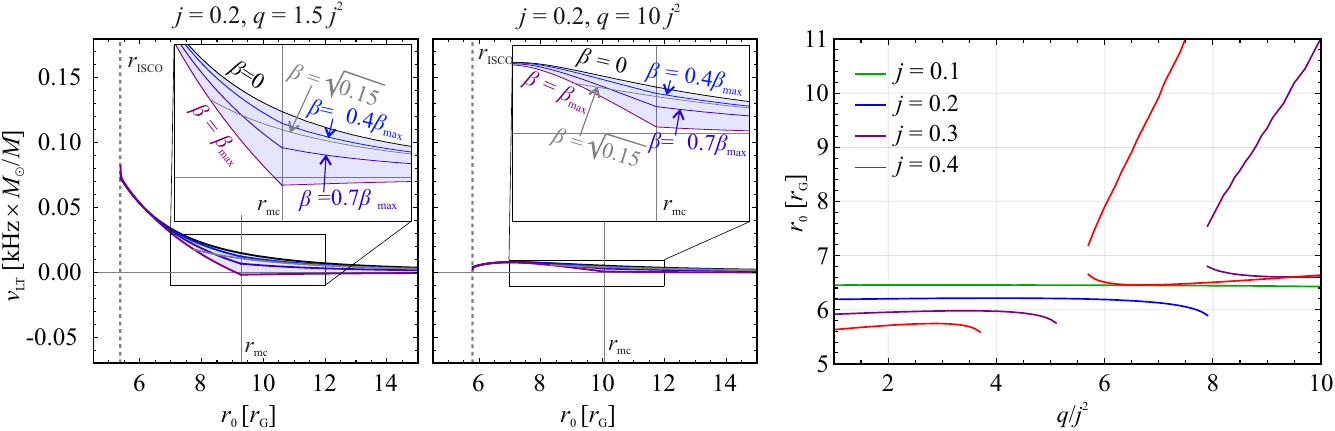}
		\caption{{\emph{Left:}} The vertical precession frequencies calculated for different $\beta$ and NSs with $j=0.2$, and two particular values of $q/j^2$, $q/j^2 = 1.5$ (extreme NS compactness), and $q/j^2 = 10$ (large NS oblateness). {\emph{Right:}} Radial coordinates corresponding to the vanishing of the correction to the vertical precession oscillation mode frequency, i.e., radii where $\nuLT^{(2)}=0$.  
  \label{figure:v_prc}}
	\end{center}
\end{figure*}

Next, for the two axisymmetric and two precession modes, we examine the quantitative and qualitative impact of the torus thickness and spacetime parameters, identify differences from the so far explored scenarios, and search for features such as possible local extrema of the frequency functions.

We also take into account the mode frequencies at various specific radii, such as the innermost stable circular orbit (ISCO), $r_\mathrm{ISCO}$. In order to do this, we consider a sequence of cusp tori with a varying torus centre position, and for a particular value of $r_0$, we mostly focus on $\beta\in[0,\,\beta_\mathrm{cusp}(r_0)]$.

\subsection{Axisymmetric modes}

The left panel of Fig.~\ref{figure:frequencies:axi} provides a comparison between the axisymmetric ($m=0$) radial mode frequencies for tori with different thickness and NSs with $j=0.2$ and different compactness. The frequencies behave qualitatively the same as the test particle's radial epicyclic frequency, being zero at ISCO, rising rather quickly with increasing radius to a distinct maximum and then slowly, almost steadily, decreasing. Because the pressure correction $\nuR^{(2)}$ is negative, the pressure forces tend to slow the epicyclic motion, and $\beta_\mathrm{cusp}$ rises with $r_0$, the maximum is located closer to the star than it would be in the case of a test particle epicyclic motion. A more detailed quantitative comparison is provided in Appendix~\ref{appendixA},  Fig.~\ref{m_0_radial}.

In a full analogy to the left panel of Fig.~\ref{figure:frequencies:axi}, the right panel provides a comparison between the axisymmetric ($m=0$) vertical mode frequencies for tori with different thickness and NSs with $j=0.2$ and different compactness. Again, the frequencies behave qualitatively like a test particle's vertical epicyclic frequency while the pressure corrections are negative. A more detailed quantitative comparison is also provided in Appendix~\ref{appendixA},  Fig.~\ref{m_0_vertical}.

\subsection{Radial precession ($m=-1$ radial mode)}
\label{Section:rprc}

The left panel of Fig.~\ref{figure:r_prc} provides a comparison between the radial precession mode frequencies for tori with different thicknesses and NSs with $j=0.2$ and different compactness. For the radii close to ISCO, the pressure corrections are negative. The maximal precession frequency is achieved at $r_0 = r_\mathrm{ISCO}$, where only infinitely slender torus with zero pressure corrections can exist. The maximum achievable frequency slightly drops with rising NS quadrupole moment as the ISCO moves to higher radii, but it is still higher than the precession rate for lower $q$ at the same (coordinate) radius.

For more distant radii, the corrections are smaller but can be positive, and at radii where they change the sign, the frequency does not depend on the torus thickness. This feature is present for both rotating and non-rotating stars. Radii relevant to this effect are shown in the right panel of Fig.~\ref{figure:r_prc}. A more detailed quantitative comparison between the radial precession mode frequencies is provided in Appendix~\ref{appendixB}, Fig.~\ref{m_1_radial}.

\subsection{Vertical precession ($m=-1$ vertical mode)}
\label{Section:vprc}

The left panel of Fig.~\ref{figure:v_prc} provides a comparison between the vertical precession mode frequencies for tori with different thicknesses and NSs with $j=0.2$ and different compactness. The frequency behaviour is even more complex than in the case of radial precession.

For compact stars, the maximal frequencies are achieved at the ISCO. For the test-particle motion and large NS oblateness, the combined influence of the oblate shape of the star and frame-dragging induce non-monotonic frequency profiles with maximum above the ISCO and emergence of a zero-frequency radius where the precession changes its orientation \citep{2013MNRAS.434.2825K,2014JPhCS.496a2016K,2016ApJ...818L..11T,2019ApJ...877...66U}. The same happens for slender and non-slender tori. At the same time, the correction to the mode frequency exhibits a behaviour similar to the case of radial precession, and radii, where the correction changes its sign exists. The relevant radii are shown in the right panel of Fig.~\ref{figure:v_prc}.

A more detailed quantitative comparison between the vertical precession mode frequencies is provided in Appendix~\ref{appendixB} (Fig.~\ref{m_1_vertical}), while an extended discussion on the mode frequency behaviour is given in Sect.~\ref{Section:EOS}.

\section{Approximate relations}
\label{Section:formulae}

The formulae for the oscillation and precession frequencies investigated in the previous section are presented in our public codes. Their explicit forms are rather lengthy and potentially too complicated for practical calculations. Inspired by our previous work \citep{2022ApJ...929...28T}, we provide their approximate versions that enable more efficient implementation. 

For each of the four modes, we postulate an approximate formula that in the limit of a slender torus provides frequencies of free test particle motion, and in the limit of a marginally overflowing (cusp) torus, fits well the results of the exact calculations of the mode frequency. The numerical coefficients in the approximate formula are found by comparing them with the results of the exact mode frequency calculations, which are performed using the least squares method.   Expressions that allow for calculations considering tori of general thickness are also included.

\subsection{Axisymmetric radial mode}

The axisymmetric radial mode frequency can be approximated as
\begin{equation}
    \label{equation:RP}
    \nuR = \nu_K \mathcal{R}\sqrt{1 - 6\mathcal{V}_0^{2/3} + 8j\mathcal{V}_0 + 57 j^2 \mathcal{V}_0^{7/3} - 3 q \mathcal{Q}}\,,
\end{equation}
where
\begin{equation}
    \label{equation:RA}
    \mathcal{V}_0 \equiv \frac{\nu_{\mathrm{K}}/\nu_{\mathrm{0}}}
    {6^{3/2} -j\nu_{\mathrm{K}}/\nu_{\mathrm{0}}
    - \tfrac{1}{2} ( j^2 - q ) ( \nu_\mathrm{K} / \nu_0 )^2 }\,,
    \quad \nu_{\mathrm{0}}=2198\frac{\mathrm{M}_\odot}{M}\,,
    \nonumber
\end{equation}
\begin{equation}
      \mathcal{Q}=\mathcal{V}_0^{4/3} \left( 1 + 19 \mathcal{V}_0 \right)\,.
\end{equation}

Note that $\nu_{\mathrm{0}}$ is the Keplerian frequency at ISCO evaluated for a non-rotating star. 

\subsubsection{Geodesic limit vs. limit of cusp torus}

\begin{figure*}[t]
	\begin{center}
		\includegraphics[width=\linewidth]{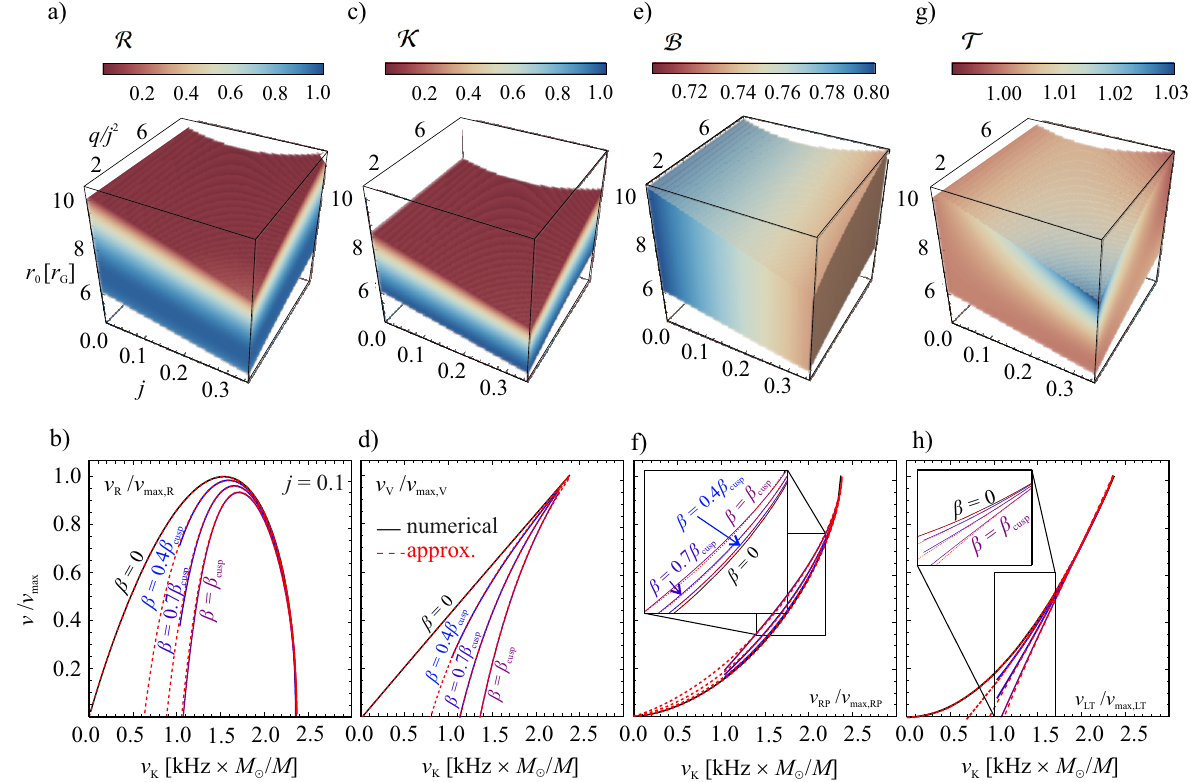} 
		\caption{Values of specific functions used in approximate relations (top)  and comparison between the exact solution and the approximate relations in the case of very slowly rotating NSs (bottom). 
  a,~b) Axisymmetric radial mode. c,~d) Axisymmetric vertical mode. e,~f) Radial precession mode. g,~h) Vertical precession mode. The frequencies are normalised with respect to their maximal values $\nu_\mathrm{max}$.
    \label{figure:cakesandfits}} 
	\end{center}
\end{figure*}

For $\mathcal{R}=1$, our formula gives the radial epicyclic frequency of a free test particle. Choosing
\begin{equation}
    \mathcal{R} = \frac{32}{33} - \frac{10 \left(\mathcal{V}_0 ^{-1} - 18.4 + 15 j -4 q\right)^2}{\nu_0 \left(1 - 1.5 j + 0.6 q\right)},
\end{equation}
we obtain a relation for the axisymmetric radial oscillations of a torus with a cusp.  The specific $\mathcal{R}$ function takes values between 0 and 1. Its dependence on spacetime parameters is illustrated in Fig.~\ref{figure:cakesandfits}a while a comparison between the (exact) mode frequencies and the approximate relation is shown in Fig.~\ref{figure:cakesandfits}b assuming very slowly rotating stars. A thorough comparison between the outcomes of exact calculations and the application of the approximate relation across the whole considered range of parameters is then provided in Appendix~\ref{appendixC} (Fig. \ref{radm0aprox}).

\subsubsection{Tori of arbitrary thickness}

To describe oscillation frequencies of tori of arbitrary thickness, we use
\begin{equation}
    \mathcal{R} = 1 - \left( \frac{\beta}{\beta_\mathrm{cusp}} \right)^2 \left( \frac{1}{33} + \frac{10 (\mathcal{V}_0 ^{-1} - 18.4 + 15 j -4 q)^2}{\nu_0 (1 - 1.5 j + 0.6 q)} \right).
\end{equation}

\subsection{Axisymmetric vertical mode}

The axisymmetric vertical mode frequency can be approximated as
\begin{equation}
    \label{equation:VA}
    \nuV = \nu_K \mathcal{K}\sqrt{1 - 4j\mathcal{V}_0 - 24 j^2 \mathcal{V}_0^{7/3} + 3 q \mathcal{V}_0^{4/3} (1 + 8 \mathcal{V}_0) }\,.
\end{equation}

\subsubsection{Geodesic limit vs. limit of cusp torus}

For $\mathcal{K}=1$, our formula provides the value of the vertical epicyclic frequency of a free test particle. Choosing 
\begin{equation}
    \mathcal{K} = 1 - \frac{14 (\mathcal{V}_0 ^{-1} - \frac{40}{3} + 10 j - 2.5 q)^2}{\nu_0 (1 - 1.7 j + 0.7 q)},
\end{equation}
we obtain an approximate relation for the frequencies of a torus with a cusp, where $\mathcal{K}\in(0,\,1)$. The dependence of the specific $\mathcal{K}$ function on spacetime parameters is illustrated in Fig.~\ref{figure:cakesandfits}, while a comparison between the (exact) mode frequencies and the approximate relation is shown in Fig.~\ref{figure:cakesandfits} assuming very slowly rotating stars. A thorough comparison between the outcomes of the exact calculations and the application of the approximate relation across the whole range of considered parameters is then provided in Appendix~\ref{appendixC}, Fig.~\ref{vertm0aprox}.

\subsubsection{Tori of arbitrary thickness}

The frequencies of tori of arbitrary thickness can be evaluated using
\begin{equation}
    \mathcal{K} = 1 - \left( \frac{\beta}{\beta_\mathrm{cusp}} \right)^2 \left( \frac{14 (\mathcal{V}_0 ^{-1} - \frac{40}{3} + 10 j - 2.5 q)^2}{\nu_0 (1 - 1.7 j + 0.7 q)} \right)\,.
\end{equation}

\subsection{Radial precession (non-axisymmetric radial mode)}

The radial precession frequency can be approximated as \citep{2022ApJ...929...28T} 
\begin{equation}
    \label{equation:RP:NONAX}
    \nuRP = \nu_{\mathrm{K}}\left[1 - \mathcal{B}\sqrt{1 - 6\mathcal{V}_0^{2/3} + 8j\mathcal{V}_0 + 57 j^2 \mathcal{V}_0^{7/3} - 3 q \mathcal{Q}}\right]\,.
\end{equation}

\subsubsection{Geodesic limit vs. limit of cusp torus}

Choosing a constant $\mathcal{B}$, $\mathcal{B}=1$, our relation with a very high accuracy provides the value of the periastron precession frequency of test particle motion around oblate neutron stars. 

When we choose
\begin{equation}
    \mathcal{B}= 0.8 - 0.2j\,, \quad
\end{equation}
the resulting relation well matches the numerical findings for the cusp torus's radial precession. The specific $\mathcal{B}$ function does not depend on the Keplerian frequency (radial coordinate $r$), unlike the case of the axisymmetric modes. 
For slowly rotating stars it takes values close to $\mathcal{B}=0.75$. The exact values are illustrated in Fig.~\ref{figure:cakesandfits}e while a comparison between the (exact) mode frequencies and the approximate relation is shown in Fig.~\ref{figure:cakesandfits}f assuming very slowly rotating stars. A thorough comparison between the outcomes of the exact calculations and the application of the approximate relation across the whole considered range of parameters is then provided in Appendix~\ref{appendixC} (Fig.~\ref{radm1aprox}).

\subsubsection{Tori of arbitrary thickness}

We may determine the relevant precession frequency for tori of arbitrary thickness by choosing
\begin{equation}
    \mathcal{B}= 1 - 0.2 (1 + j) \left( \frac{\beta}{\beta_\mathrm{cusp}} \right)^2\,.
\end{equation}

\subsection{Vertical precession}

The vertical precession frequency can be approximated as
\begin{equation}
    \label{equation:LT}
    \nuLT = \nu_K \mathcal{T}\sqrt{1 - 4j\mathcal{V}_0 - 24 j^2 \mathcal{V}_0^{7/3} + 3 q \mathcal{V}_0^{4/3} (1 + 8 \mathcal{V}_0) }\,.
\end{equation}

\subsubsection{Geodesic limit vs. limit of cusp torus}

Choosing $\mathcal{T} = 1$, our formula provides the value of the vertical precession frequency for a free test particle. Selecting 

\begin{align}
    \mathcal{T} = 1 - \frac{5 j}{\nu_0} \left[ 1 - \frac{2 ( \mathcal{V}_0^{-2/3} - 6.1 + 2.5 j )^2}{1 - 2 j - 2 j^2 + q}  \right]\\ \nonumber 
                    + \frac{9 }{\nu_0} ( q - j^2 ) ( \mathcal{V}_0^{-2/3} - 7.2 ),
\end{align}

we can describe the vertical precession of a cusp torus. The specific $\mathcal{T}$ function takes values very close to 1.  The exact values are illustrated in Fig.~\ref{figure:cakesandfits}g while a comparison between the (exact) mode frequencies and the approximate relation is shown in Fig.~\ref{figure:cakesandfits}h assuming very slowly rotating stars. A thorough comparison between the outcomes of the exact calculations and the application of the approximate relation across the whole considered range of parameters is then provided in Appendix~\ref{appendixC} (Fig.~\ref{vertm1aprox}).

\subsubsection{Tori of arbitrary thickness}

To describe a torus of arbitrary thickness, we use 
\begin{align}
    \mathcal{T} = 1 - \left( \frac{\beta}{\beta_\mathrm{cusp}} \right)^2 
                \left\{ \frac{5 j}{\nu_0} \left[ 1 - \frac{2 ( \mathcal{V}_0^{-2/3} - 6.1 + 2.5 j )^2}{1 - 2 j - 2 j^2 + q}  \right]   \right. \nonumber \\
                   \left. + \frac{9}{\nu_0} ( q - j^2 ) ( \mathcal{V}_0^{-2/3} - 7.2 ) \right\} \,.
\end{align}

\section{Discussion and conclusions}
\label{Section:conclusions}

The impact of the torus thickness, in most cases, leads to relatively high negative corrections to the modes' frequencies, which should be reflected within possible astrophysical applications. Furthermore, as found in Sects.~\ref{Section:rprc} and \ref{Section:vprc}, for the precession modes, the pressure corrections can be positive within a limited range of radii because the special radii exist where the precession frequency does not depend on the torus thickness. This can have observable consequences. Nevertheless, since we are constrained by the adopted perturbative approach and the effect manifests itself on a relatively large radius where the scatter between the frequencies calculated for different tori is generally very small, it will be useful to verify its presence by a suitable future numerical solution of Eq.~(\ref{papa-L}).

\begin{figure*}[t]
	\begin{center}
		\includegraphics[width=1.\linewidth]{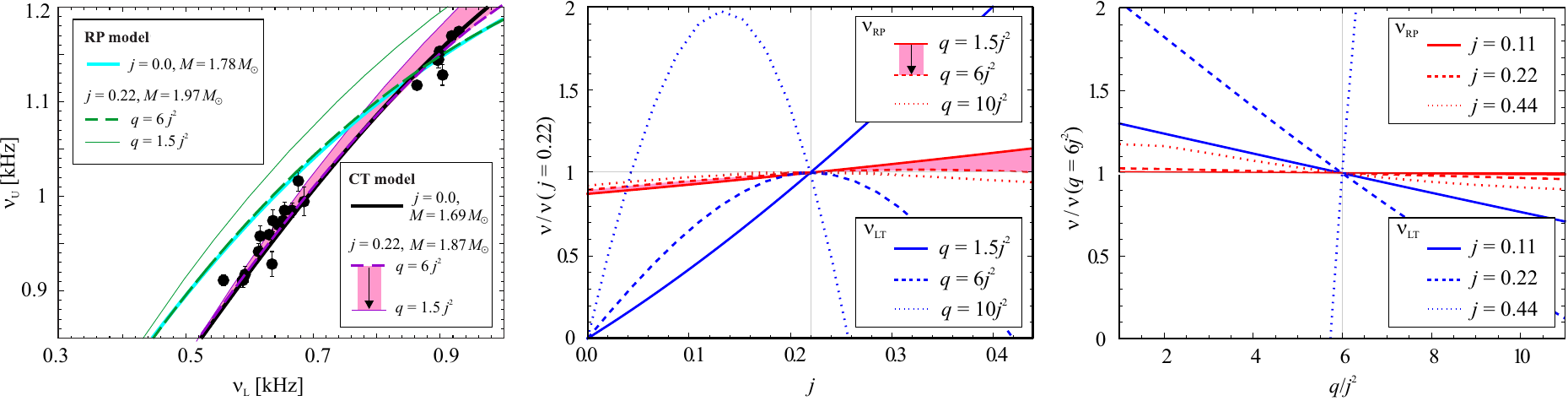}
		\caption{{\emph{Left:}} Correlations between the lower and upper QPO frequencies, $\nu_{\mathrm{L}}$ and $\nu_{\mathrm{U}}$, predicted by the CT  and RP models, compared to the observed data from 4U~1636-53 atoll source.  The best fits predicted for non-rotating NSs are denoted by thick solid lines, while the best fits for moderately rotating NSs with $j=0.22$ and $q=6j^2$ are denoted by dashed lines. The thin solid lines are then drawn for these values of $M$ and $j$ but different values of $q$ ($q=1.5j^2$). For the CT model, the light-coloured region emphasises the change of the frequencies predicted by the model with changing $q$. \emph{Middle:} Relative change of the cusp torus precession frequencies with angular momentum at the radius where $\nuRP=0.75\nu_{\mathrm{K}}$. Within the RP and CT models, this radius corresponds to the bottom left group of data points shown in the left panel. The light-coloured region then emphasises the corresponding change of $\nuRP$.  {\emph{Right}}: Relative change of precession frequencies with the quadrupole parameter. Frequencies are calculated at the same radius as considered within the middle panel.
		\label{figure:impact}
		}
	\end{center}
\end{figure*}

\subsection{Influence of NS parameters}
 
Both the NS rotation and rotationally induced oblateness affecting the angular momentum and quadrupole moment of the NS have significant effects on the frequencies of the investigated tori oscillations. The impact of the quadratic relations behind the nature of the modes is well illustrated in terms of quadrupole parameter $q/j^2$ used in our study, which for a NS with a given EoS and mass does not depend on the NS rotational frequency \citep{1968ApJ...153..807H, 2019ApJ...877...66U} . This parameter effectively ranging from $q/j^2\sim1.5$ to $q/j^2\sim10$ determines whether the influence of the $t$-$\phi$ component of the metric tensor and related frame dragging effects dominate (low values of the parameter corresponding to Kerr like behaviour) or are overwhelmed by the other components (high values of the parameter corresponding to the effective suppression of frame-dragging effects by the NS oblateness).

This interaction between the relativistic frame dragging effects associated with the angular momentum and the effects associated with the quadrupole moment, which would also arise in Newtonian physics, determines the resulting behaviour of the frequencies of the studied modes. For low values of $q/j^2$, the maximal frequencies of the modes increase with $j$ within the considered range. For large values of $q/j^2$, these frequencies increase only minimally or even decrease with increasing $j$.

\subsection{Consequences for models of NS variability - HF QPOs}
\label{Section:QPOs}

\citet{2016MNRAS.457L..19T} and \cite{2022ApJ...929...28T}  have introduced a QPO model that identifies the twin peak QPO frequencies with the Keplerian frequency, $\nu_{\mathrm{K}}=\Omega_0/2\pi$, and the radial precession frequency, $\nuRP$, of the fluid in the cusp-torus configuration (CT model). In the left panel of Fig.~\ref{figure:impact}, we compare the best fits to the frequencies observed in the 4U~1636-53 atoll source \citep[the data are taken from][]{2006MNRAS.370.1140B,2009A&A...497..661T} obtained using the CT model to the best fits based on the relativistic precession model \citep[RP model,][]{1999PhRvL..82...17S,1999ApJ...524L..63S}.

From Fig.~\ref{figure:impact}, we can see that the RP model, which deals with geodesic precession, provides less promising fits of the data than the CT model. The effects associated with the NS rotation do not bring significant improvement in comparison to the non-rotating case and rather change the best-fitting mass value because the fitting-parameter space is almost degenerate \citep[][]{2016MNRAS.457L..19T,2022ApJ...929...28T}. Moreover, when we restrict ourselves to values of the Hartle-Thorne spacetime parameters that are consistent with up-to-date models of NSs, or impose even tighter constraints, such as $j<0.3$, $M<2.3M_\sun$, $q/j^2<10$, no conceivable smooth monotonic curve can reproduce the data in a significantly better way compared to the CT model.

\subsection{Consequences for models of NS variability - LF QPOs}
\label{Section:QPOsLF}

In principle, the relations derived here, including those determining the vertical precession frequency, can be used in an analogous way to put constraints on models of low-frequency QPOs providing that they are associated with the vertical precession. Following this research direction, initiated in the context of the RP model, can be very promising. In the middle and right panels of Fig.~\ref{figure:impact}, we consider a particular orbital radius where $\nuRP=0.75\,\nu_{\mathrm{K}}$. Within the RP and CT models, this radius roughly corresponds to a large part of the documented QPO detections ($r=6.75\rg$ when $j=0$). These panels clearly illustrate that the vertical precession frequency of the cusp torus is more sensitive to angular momentum and quadrupole moment than the radial one. This is caused by the strong vertical change of the gravitational field in the vicinity of an oblate NS, which already affects the test particle motion \citep[][]{1999ApJ...513..827M, 2013MNRAS.434.2825K} and then applies to the fluid motion in tori of any thickness.

In order to explore detailed consequences of the behaviour of $\nuLT$ in relation to QPO models and diagnostics of NS parameters, further investigation is needed. As shown in the middle panel of Fig.~\ref{figure:impact}, the dependence of $\nuLT$ on the NS spin can (for both test particle and fluid motion) exhibit a clear maximum, which can be of astrophysical importance.

\begin{figure*}[t]
	\begin{center}
		\includegraphics[width=.95\linewidth]{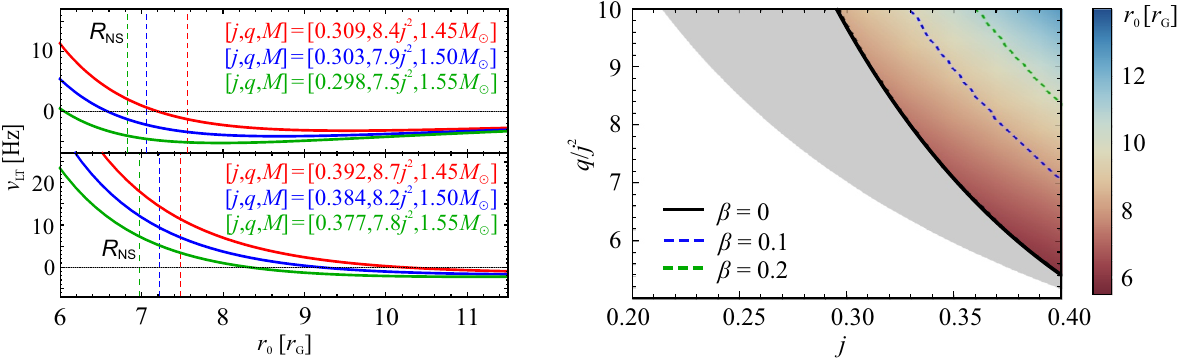}
		\caption{{\emph{Left:}} Profiles of the slender torus vertical precession frequency for a particular choice of spacetime parameter's values, along with the related NS radii calculated using the {L} NS EoS \citep[][]{L} and denoted by vertical dashed lines. The top (bottom) panel corresponds to a rotational frequency of 400Hz (500Hz). {\emph{Right:}} The range of the Hartle-Thorne spacetime parameters for which the vertical precession may change its orientation.  The thick black line marked as $\beta = 0$ indicates the limiting NS parameters relevant for the case of a slender torus, which are the same as for the geodesic (thin disc) case explored by \cite{2016ApJ...818L..11T}. The shaded region below the $\beta = 0$ line determines the area where this effect may occur, but the corresponding tori overlaps with the star's surface. The coloured region above the $\beta = 0$ line corresponds to the tori that do not overlap with the star, while the colour scale indicates the radii of their centres for the slender torus case. In a full analogy, the lines for $\beta = 0.1$ and $\beta = 0.2$ show the limiting parameters where the torus of the corresponding thickness would touch the star's surface. In these cases, the corresponding $r_0$ colourmaps would be slightly different but almost identical to the $\beta = 0$ map shown in this figure. \label{figure:EoSLT}}
  
	\end{center}
\end{figure*}

\subsection{Impact of NS EoS}
\label{Section:EOS}

So far we focused on examining the effects of spacetime parameters on the investigated oscillations while taking into account a range of these parameters associated with present models of NSs. As follows from the results of Paper~I, particular astrophysical applications of our formulae should be rather sensitive to the NS EoS, most importantly to the specific position of the NS surface. 

 An example of the impact of the NS EoS choice can be demonstrated by the effect of the vertical precession's change of orientation discussed in Sect.~\ref{Section:vprc}. In the left panel of Fig.~\ref{figure:EoSLT}, we illustrate profiles of the vertical precession frequency for a particular choice of spacetime parameter's values along with several related NS radii which are calculated using the {L} NS EoS introduced by \citet{L}. From the figure, we can see that the investigated effect occurs at the radii that are not necessarily above the NS surface. We use the {L} NS EoS, as it is rather extremely stiff and therefore well allows us to illustrate the effect of a large quadrupole moment, but the effect can be relevant for many currently considered EoS.
 
 In the right panel of Fig.~\ref{figure:EoSLT}, we show the full range of NS's angular momentum and quadrupole moment, which corresponds to the presence of the effect. Utilising the recently identified universal relations \citep[][]{2013PhRvD..88b3007M,2013MNRAS.433.1903U,2013PhRvD..88b3009Y,2013Sci...341..365Y,2015MNRAS.454.4066P,2017MNRAS.470L..54R}, we denote the coloured area which determines the range limited by the condition that the inner edge of the torus should be located above the NS surface.  Clearly, this range is smaller than the full range given by the general consideration of the Hartle-Thorne metric alone.

\subsection{Summary and overall implications}

In this work, we studied the epicyclic oscillations of fluid tori in the Hartle-Thorne geometry, paying special attention to the two axisymmetric and two precession modes. We focused on the case of discs with constant angular momentum distribution and polytropic index $n=3$.\footnote{In Appendix~\ref{appendixD}, where we consider different values of $n$, we show that the effect of the choice of $n$ is rather very small, except in the case of the vertical axisymmetric mode.} In view of the above discussion, we can draw the conclusion that the consideration of the quadrupole moment induced by the NS rotation likely has a strong impact on various astrophysical phenomena involving oscillations of the accreted fluid, such as the observed high and low-frequency quasi-periodic oscillations. The formulae derived in our work can help quantify this impact. While here we provided all the formulae required for a further examination of the underlying parameter space, particular astrophysical applications that pose the potential to restrict the dense-matter EoS will need to be carried out, focusing on specific models of NSs.

\begin{acknowledgements}
     We thank Marek Abramowicz, Omer Blaes, and W{\l}odek Klu{\'z}niak for the valuable discussions. We wish to express our gratitude to the University of California in Santa Barbara for hosting us while this study was being initiated. We also thank the reviewer for valuable comments and suggestions, which significantly helped to improve the paper. We acknowledge the Czech Science Foundation (GA\v{C}R) grant No.~21-06825X and the INTER-EXCELLENCE project No. $\mathrm{LTI17018}$, and the PRODEX program of the European Space Agency (ref.\ 4000132152). The INTER-EXCELLENCE project No. LTT17003 is acknowledged by KK, and the INTER-EXCELLENCE project No. LTC18058 is acknowledged by MM and MU. VK acknowledges the Research Infrastructure LM2023047 of the Czech Ministry of Education, Youth and Sports. OS acknowledges funding by the Deutsche Forschungsgemeinschaft (DFG, German Research Foundation) under Germany's Excellence Strategy – EXC 2094 – 390783311. We furthermore acknowledge the support provided by the internal grants of Silesian University, $\mathrm{SGS/31/2023}$ and $\mathrm{SGS/25/2024}$.  
\end{acknowledgements}
 
\bibliographystyle{aa}
\bibliography{mat}

\begin{appendix} 
\onecolumn
\section{Frequencies of axisymmetric modes -- detailed view}
\label{appendixA}

 \begin{figure*}[h!t]
	\begin{center}
		\includegraphics[width=0.92\linewidth]{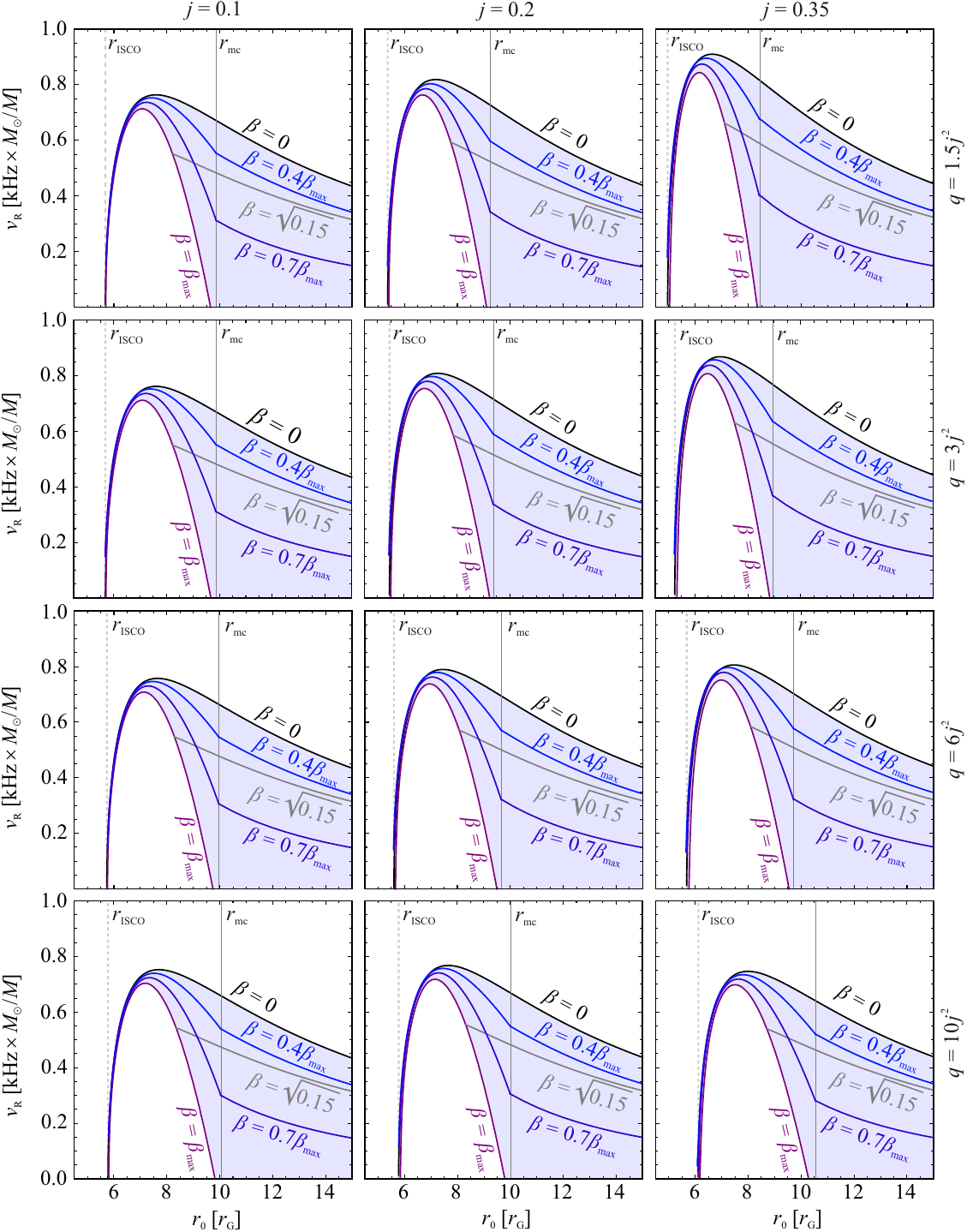}
		\caption{ 
  The axisymmetric radial oscillation frequencies of tori with different $\beta$ plotted for the NS dimensionless angular momentum ($j=0.1$, $j=0.2$ and $j=0.35$) and its dimensionless quadrupole parameter ($q/j^2 = 1.5$ -- extreme compactness, $q/j^2 = 3$ -- small oblateness, $q/j^2 = 6$ -- moderate oblateness and $q/j^2 = 10$ -- large oblateness).
		\label{m_0_radial}
		}
	\end{center}
\end{figure*}

\begin{figure*}[h!t]
	\begin{center}
		\includegraphics[width=0.93\linewidth]{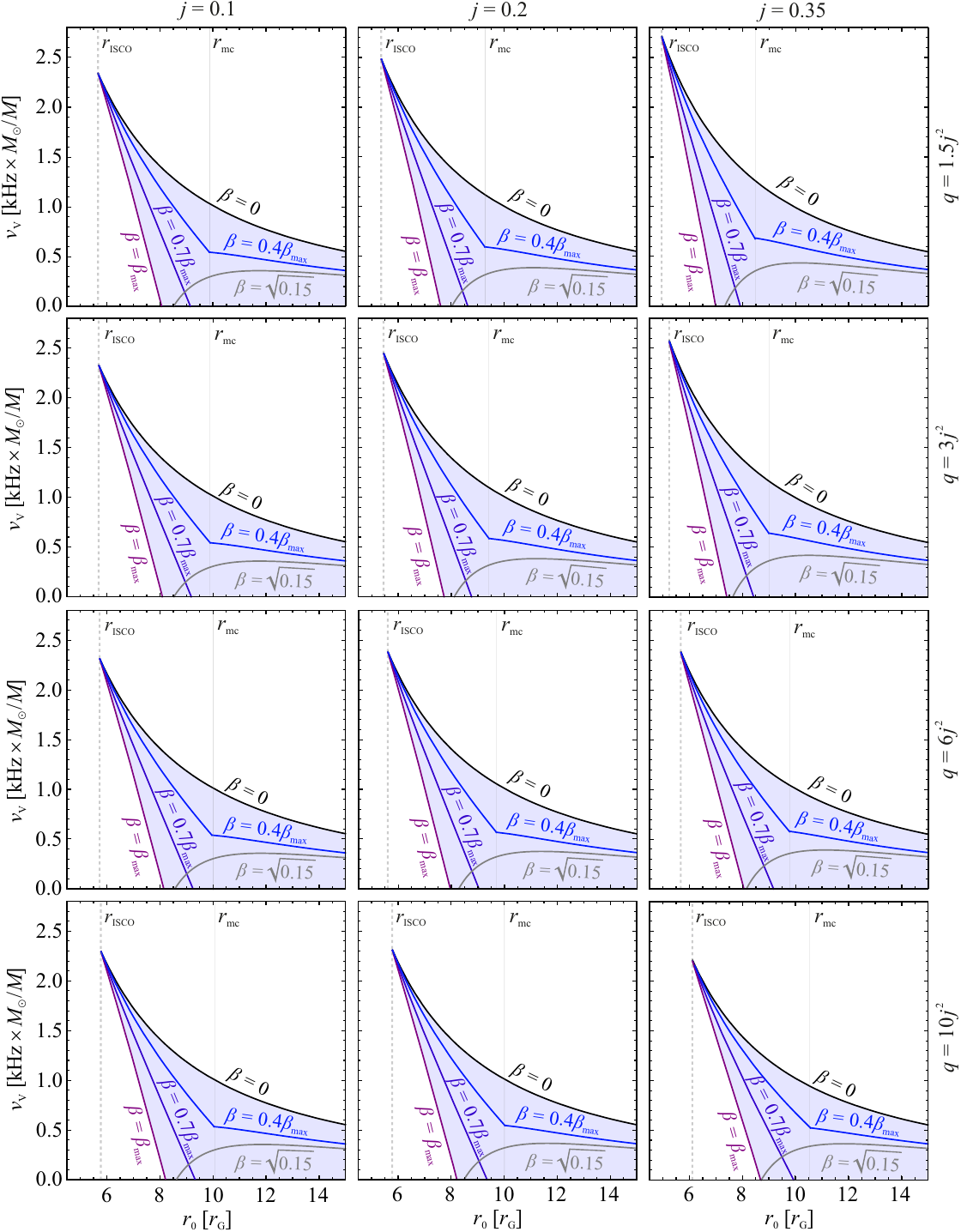}
		\caption{ 
  The axisymmetric vertical epicyclic frequencies of tori with different $\beta$ plotted for different values of NS dimensionless angular momentum ($j=0.1$, $j=0.2$ and $j=0.35$) and the quadrupole parameter ($q/j^2 = 1.5$ -- extreme compactness, $q/j^2 = 3$ -- small oblateness, $q/j^2 = 6$ -- moderate oblateness and $q/j^2 = 10$ -- large oblateness).
		\label{m_0_vertical}
		}
	\end{center}
\end{figure*}

\section{Frequencies of non-axisymmetric modes -- detailed view}
\label{appendixB}

\begin{figure*}[h!t]
	\begin{center}
		\includegraphics[width=0.92\linewidth]{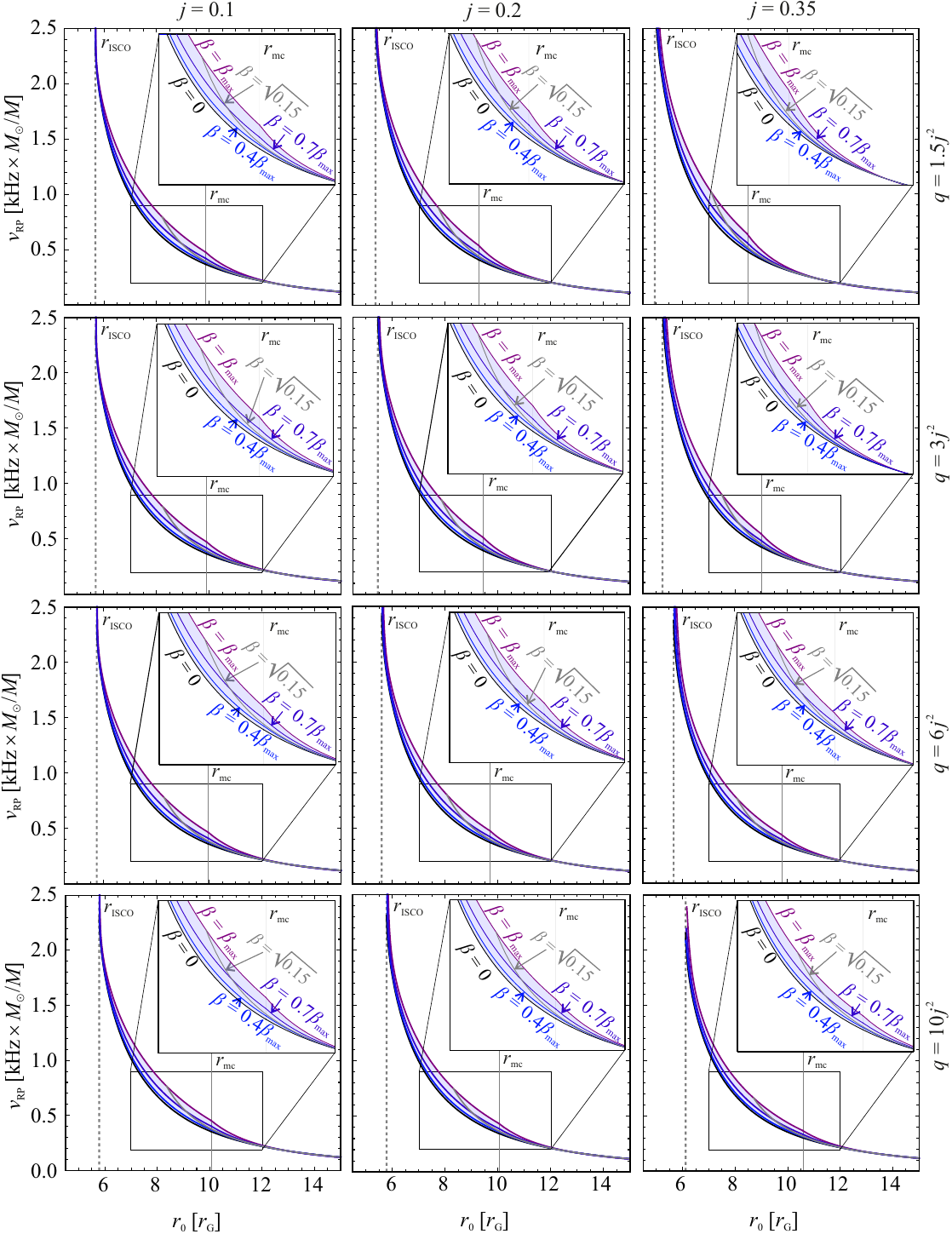}
		\caption{Radial precession frequencies of tori with different $\beta$ plotted for different values of NS angular momentum ($j=0.1$, $j=0.2$ and $j=0.35$) and the quadrupole parameter ($q/j^2 = 1.5$ -- extreme compactness, $q/j^2 = 3$ -- small oblateness, $q/j^2 = 6$ -- moderate oblateness and $q/j^2 = 10$ -- large oblateness).
		\label{m_1_radial}
		}
	\end{center}
\end{figure*}

\begin{figure*}[h!t]
	\begin{center}
		\includegraphics[width=0.92\linewidth]{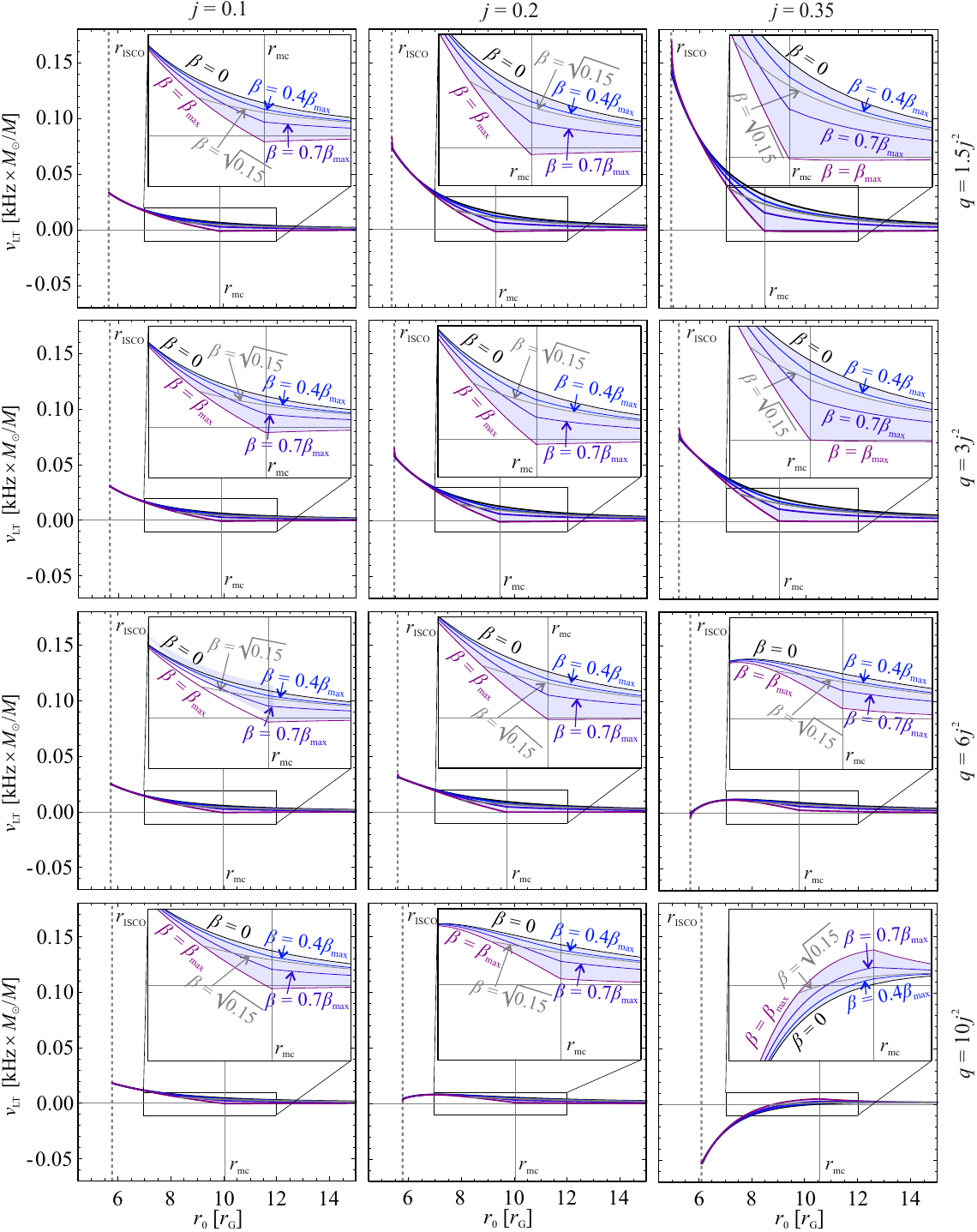}
		\caption{Vertical precession frequencies of tori with different $\beta$ for NS angular momentum ($j=0.1$, $j=0.2$ and $j=0.35$) and its dimensionless quadrupole parameter ($q/j^2 = 1.5$ -- extreme compactness, $q/j^2 = 3$ -- small oblateness, $q/j^2 = 6$ -- moderate oblateness and $q/j^2 = 10$ -- large oblateness).
		\label{m_1_vertical}
		}
	\end{center}
\end{figure*}

\newpage
\section{Approximative relations -- detailed view}
\label{appendixC}

\begin{figure*}[hb]
	\begin{center}
		\includegraphics[width=0.93\linewidth]{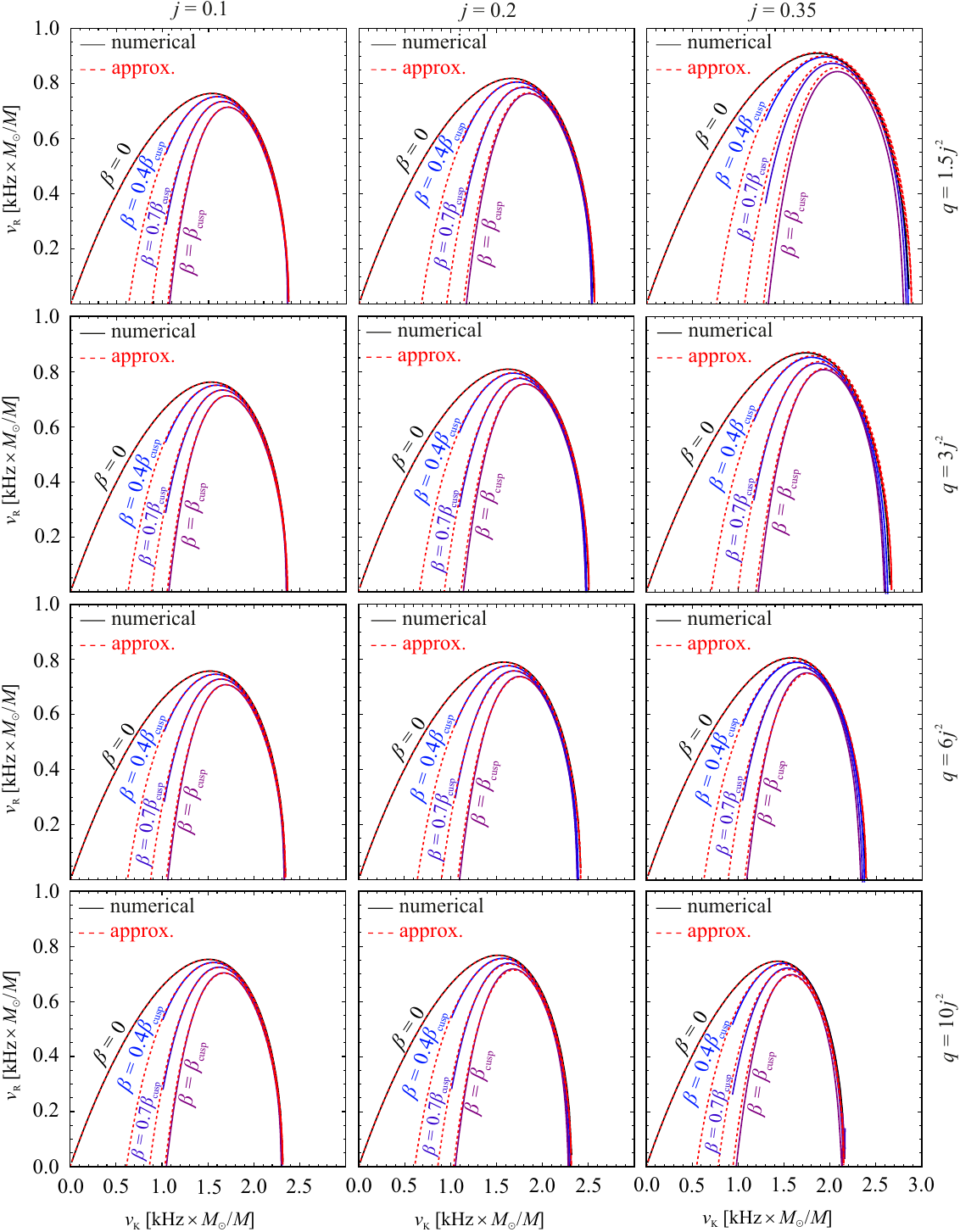}
		\caption{Comparison of the radial epicyclic frequency of tori with different thicknesses (black) and the approximation relations (red). \label{radm0aprox}}
	\end{center}
\end{figure*}

\begin{figure*}[bt]
	\begin{center}
		\includegraphics[width=0.93\linewidth]{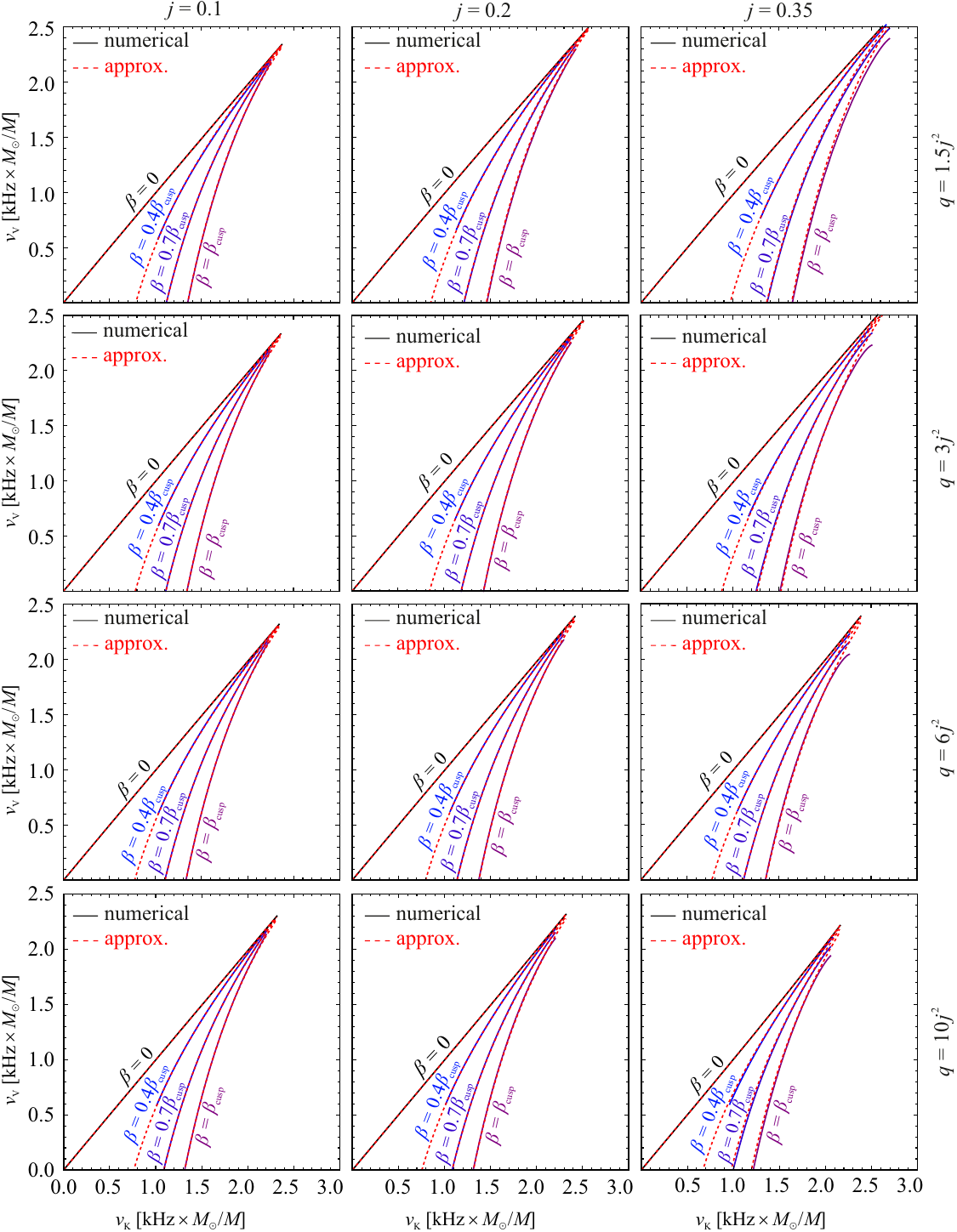}
		\caption{Comparison of the vertical epicyclic frequency of tori with different thicknesses (black) and the approximation relations (red). \label{vertm0aprox}}
	\end{center}
\end{figure*}

\begin{figure*}[t]
	\begin{center}
		\includegraphics[width=0.93\linewidth]{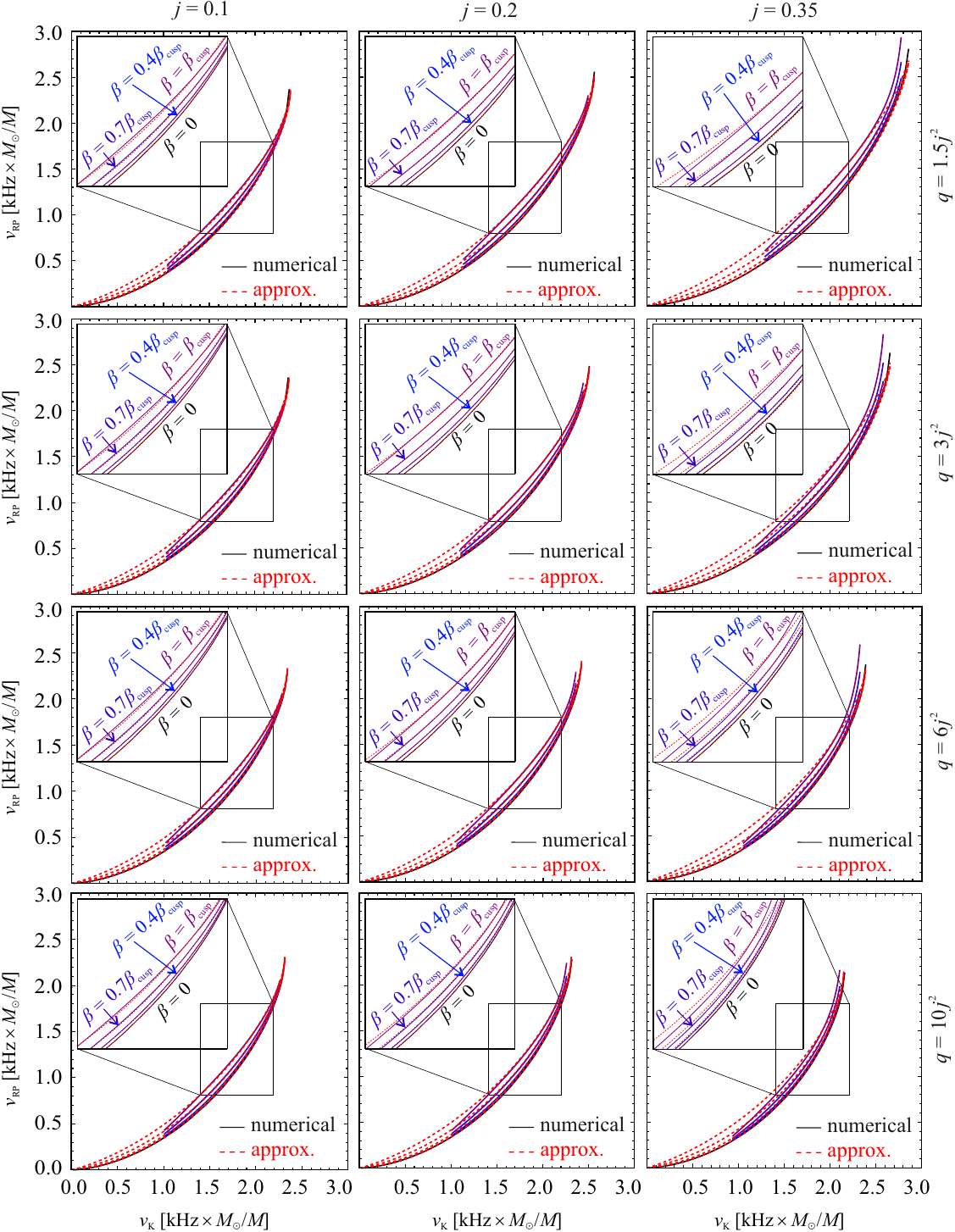}
		\caption{Comparison of the radial precession frequency of a free particle (black) and torus (blue) to the approximation relations (red). \label{radm1aprox}}
	\end{center}
\end{figure*}

\begin{figure*}[t]
	\begin{center}
		\includegraphics[width=0.93\linewidth]{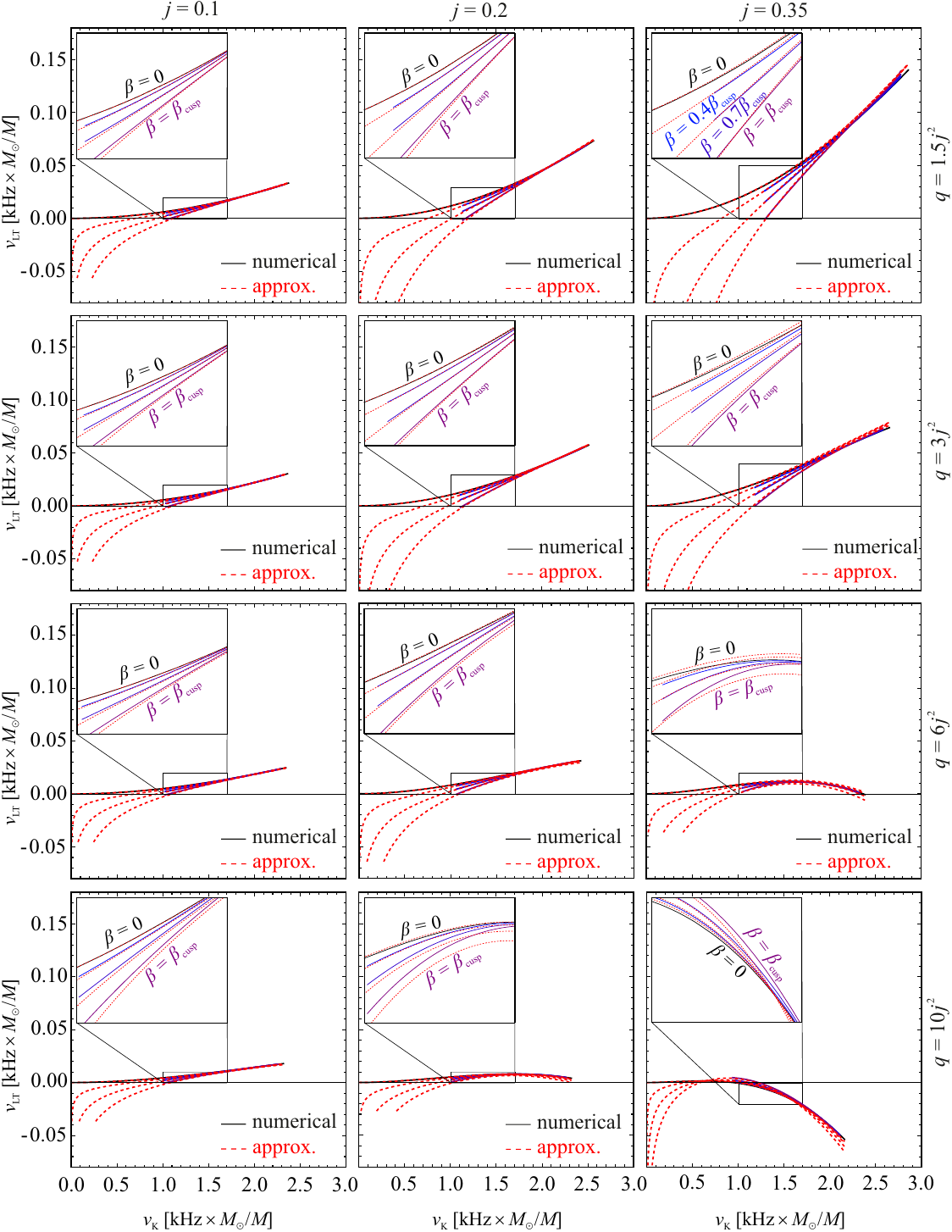}
		\caption{Comparison of the vertical precession frequency of a free particle (black) and torus (blue) to the approximation relations (red). \label{vertm1aprox}}
	\end{center}
\end{figure*}

\clearpage
\section{Impact of a change in the polytropic index}
\label{appendixD}

\begin{figure*}[h!]
	\begin{center}
		\includegraphics[width=0.88\linewidth]{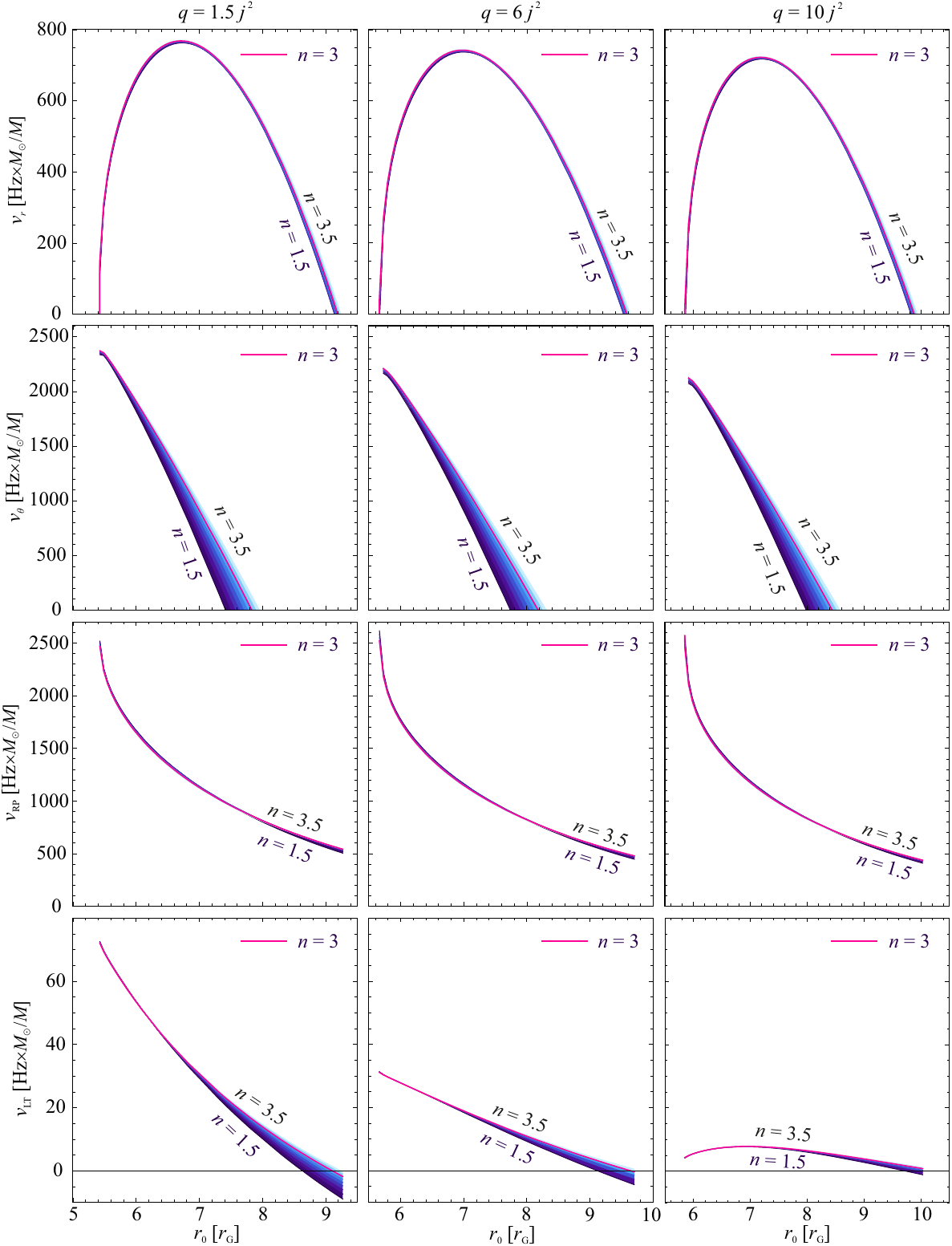}
		\caption{Frequencies of axisymmetric oscillations and precession frequencies of cusp tori as a function of the polytropic index for $j=0.2$. Frequency values corresponding to the range $n\in[1.5,\,3.5]$ are scaled in shades of blue. The value of $n=3$ is emphasised by purple. \label{diffn}}
	\end{center}
\end{figure*}

\end{appendix}

\end{document}